\documentclass[fleqn,usenatbib]{mnras}

\usepackage{newtxtext,newtxmath}

\usepackage[T1]{fontenc}

\DeclareRobustCommand{\VAN}[3]{#2}
\let\VANthebibliography\thebibliography
\def\thebibliography{\DeclareRobustCommand{\VAN}[3]{##3}\VANthebibliography}

\usepackage{graphicx}	
\usepackage{amsmath}	

\title[Galaxy Morphology through an Unsupervised-Supervised Hybrid Approach]{Unveiling Galaxy Morphology through an Unsupervised-Supervised Hybrid Approach}

\author[Kolesnikov et al. 2023]{
I. Kolesnikov,$^{1,4}$\thanks{E-mail: igor.s.kolesnikov@gmail.com}
V. M. Sampaio,$^{2,3}$
R. R. de Carvalho,$^{3}$
C. Conselice,$^{4}$
S. B. Rembold,$^{5}$
C. L. Mendes,$^{1}$
\newauthor Rosa, R. R., $^{1}$
\\
$^{1}$ Instituto Nacional de Pesquisas Espaciais, São José dos Campos, 12227-010, SP, Brazil\\
$^{2}$ School of Physics and Astronomy, University of Nottingham, University Park, Nottingham NG7 2RD, UK\\
$^{3}$ NAT - Universidade Cidade de São Paulo, 01506-000, SP, Brazil\\
$^{4}$ Jodrell Bank Centre for Astrophysics, University of Manchester, Oxford Road, Manchester M13 9PL, UK\\
$^{5}$Universidade Federal de Santa Maria, Santa Maria, RS, 97105-900, Brazil
}

\date{Accepted XXX. Received YYY; in original form ZZZ}

\pubyear{2023}

\begin{document}
\label{firstpage}
\pagerange{\pageref{firstpage}--\pageref{lastpage}}
\maketitle

\begin{abstract}
Galaxy morphology offers significant insights into the evolutionary pathways and underlying physics of galaxies. As astronomical data grows
with surveys such as Euclid and Vera C. Rubin
, there is a need for tools to classify and analyze the vast numbers of galaxies that will be observed. 
In this work, we introduce a novel classification technique blending unsupervised clustering based on morphological metrics with the scalability of supervised Convolutional Neural Networks. We delve into a comparative analysis between the well-known \texttt{CAS} (Concentration, Asymmetry, and Smoothness) metrics and our newly proposed \texttt{EGG} (Entropy, Gini, and Gradient Pattern Analysis). Our choice of the \texttt{EGG} system stems from its separation-oriented metrics, maximizing morphological class contrast. We leverage relationships between metrics and morphological classes, leading to an internal agreement between unsupervised clustering and supervised classification. Applying our methodology to the Sloan Digital Sky Survey data, we obtain $\sim$95\% of Overall Accuracy of purely unsupervised classification and when we replicate \texttt{T-Type} and visually classified galaxy catalogs with accuracy of $\sim$88\% and $\sim$89\% respectively, illustrating the method's practicality. Furthermore, the application to Hubble Space Telescope data heralds the potential for unsupervised exploration of a higher redshift range. A notable achievement is our $\sim$95\% accuracy in unsupervised classification, a result that rivals when juxtaposed with Traditional Machine Learning and closely trails when compared to Deep Learning benchmarks.
\end{abstract}



\section{Introduction}

Galaxy morphology has been a cornerstone of extragalactic astronomy since the early 20th century. The pioneering work of \citet{hubble1926,hubble1936realm} introduced a classification system that divided galaxies into two broad categories: those with a dominant bulge component, known as early-type galaxies (ETGs), and those with a significant disk component, referred to as late-type or spiral galaxies (LTGs). This classification scheme has since been pivotal in understanding the large-scale structure of the universe.

Morphological classification is not merely a taxonomic exercise; it is deeply intertwined with the physical processes governing galaxies. Additional information, such as spectral properties, has shown that the morphology and physical properties of galaxies are intimately related (e.g., \citealt{nair2010fraction}). In the local universe, most 
of the 
elliptical galaxies exhibit redder colors and older stellar populations compared to spiral galaxies. The latter are predominantly gas-rich, star-forming systems with high rotation velocities \citep{roberts1994, blanton2009, pozzetti2010, kauffmann2003dependence}. Such distinctions in stellar properties underscore the importance of morphology in understanding the underlying astrophysical processes, such as star formation and galaxy evolution. Moreover, the morphology of a galaxy is closely related to its evolutionary history and the environment it resides in. 
For instance, \citet{Dressler} establishes a relation between galaxy morphology and galaxy local density.
Specifically, ETG (also known as Elliptical) dominate the central parts of clusters of galaxies (high density and high velocity dispersion), while LTG ( also known as Spiral) are primarily found in the outskirts of these systems (low density and lower velocity dispersion). Understanding the so-called “morphology-density relation” is of paramount importance in solving the puzzle that galaxy evolution represents.

In the current era of large astronomical surveys, the challenge lies in robustly classifying the vast number of galaxies there is. Traditional visual classifications are time-consuming and challenging to scale for extensive datasets. Projects such as Galaxy Zoo 1 (GZ1)\citep{lintott2011galaxy} and Zooniverse \citep{simpson2014zooniverse}, rooted in citizen science, exist and serve as foundational ground-truth catalogs for numerous studies. However, these are fundamentally based on the human eye, which might introduce biases or overlook specific details. The human eye can serve as a reliable tool for galaxy morphology assessment, depending on the quality and nature of the available data. Visual inspection by astronomers and researchers offers a unique advantage, capturing intricate details and nuances that automated algorithms might miss. Given high-resolution images and well-processed data, human observers can discern subtle features like spiral arms, bars, and irregularities that typify various galaxy types. Moreover, the adaptability of the human eye allows for the recognition of unanticipated morphological anomalies or transitional forms not catered for in predefined classification schemes. 
Nonetheless, the reliability of visual assessment depends on the observer's expertise, potential biases, and the inherent limitations of visual perception.
Integrating visual inspection with quantitative analysis, bolstered by machine learning techniques, can therefore establish a balance. This harmonizes the discerning power of the human eye with the consistency and objectivity of computational methods, further enriching our understanding of galaxy morphology. Citizen science projects have significantly advanced the classification of millions of galaxies through visual inspection. For instance, the GZ1 catalog encompasses nearly one million galaxies. While it boasts a high level of accuracy for selected galaxies, its granularity is limited due to its binary classification approach \citep{lintott2011galaxy}. With the advent of machine learning and deep learning methodologies, automated classifications have become more accurate and efficient \citep{ferrari2015morfometryka, dominguez2018improving, barchi2020machine, khalifa2017deep, primack2018deep, khan2019deep, tohill2021quantifying, walmsley2022galaxy, cheng2023lessons}. This progression has led to more detailed morphological studies, such as the one by \citet{dominguez2018improving}. This work builds upon the GZ1 catalog by introducing a \texttt{T-Type} value for each galaxy. The \texttt{T-Type} acts as a nuanced numerical descriptor and can be mapped to morphological types: ETGs have \texttt{T-Type} \( \leq 0 \), while LTGs have \texttt{T-Type} \( > 0 \) \citep{de1963revised}. 
This addition greatly enhances the granularity and precision of the catalog. 
Automated techniques rooted in visual morphology have already demonstrated significant promise in galaxy morphology classifications \citep{ferrari2015morfometryka, dominguez2018improving, barchi2020machine, khalifa2017deep, primack2018deep, khan2019deep}. They can process extensive datasets efficiently and derive meaningful patterns and insights. Combining visually classified galaxies with machine learning methodologies presents a promising avenue for delving into galaxy morphology. However, even with these sophisticated tools, a sizable fraction of the gathered data remains either unprocessed or unanalyzed, or may contain biases stemming from human judgment.

The growing interest in developing novel methods to extract meaningful information from data stems not only from the sheer volume of this data but also from its escalating complexity. Instruments like the Sloan Digital Sky Survey (SDSS) have meticulously catalogued vast regions of the sky, generating terabytes of data and capturing images of millions of celestial entities \citep{eisenstein2011}. As we transition into a new epoch of observational astronomy, missions such as Euclid \citep{euclid2011} and the Large Synoptic Survey Telescope (LSST) are set to eclipse these achievements, amassing petabytes of data and imaging billions of galaxies \citep{racca2016, ivezic2019}. While this wealth of data holds the potential to revolutionize our understanding of the universe, it also presents significant challenges. The sheer volume of data makes manual analysis, such as visual classifications of galaxy morphologies, impractical if not impossible. Moreover, the intricate details and subtle patterns in the data, which could hold clues to novel astrophysical phenomena, can easily be overlooked by visual inspection.

While galaxy morphology is often visually striking, it necessitates quantitative metrics to systematically investigate and categorize the vast array of galactic structures. One of the most used methods to systematize observed morphological varieties hinges on the combination of Concentration (\texttt{C}), Asymmetry (\texttt{A}), and Smoothness (\texttt{S}) metrics, often referred to as the \texttt{CAS} system \citep{conselice2003relationship}. These metrics have proven instrumental in probing the physical attributes of galaxies. For example, \texttt{C} correlates with the Bulge-to-Total light ratio, a critical distinguisher between spirals and ellipticals \citep{conselice2003relationship}. Moreover, \citet{graham2001correlation} demonstrated a potential correlation of \texttt{C} like velocity dispersion, galaxy size, luminosity, and black hole mass. \citet{conselice2000asymmetry} and \citet{ conselice2003relationship} 
suggests
that \texttt{A} serves as a reliable morphological marker for galaxies undergoing interactions and mergers
nonetheless it's efficiency is yet to be tested in extending morphological analysis to higher redshifts.
Further, \citet{conselice2003relationship} explored how the \texttt{S} metric can shed light on the clumpiness of galaxies, thereby revealing insights into their star formation history. As observed, the \texttt{S} metric adeptly distinguishes ETGs from LTGs. Beyond merely separating ellipticals and spirals these indicators can be assessed across varying environments and redshifts, thereby illuminating the evolutionary pathways of galaxies.

Beyond the \texttt{CAS} system, additional metrics have emerged that effectively discriminate between ETGs and LTGs. For instance, there is the Gini coefficient and M20 \citep{lotz2004new}.  
Its values range from 0, indicating a perfectly uniform distribution, to 1, indicative of maximal inequality \citep{abraham2003new}. Meanwhile, \texttt{M20} captures the spatial extent of the brightest 20\% of pixels within a galaxy \citep{lisker2008gini,lotz2004new}. More contemporaneously, \citet{ferrari2015morfometryka} have introduced entropy (\texttt{E}), which gauges the heterogeneity of light distribution 
measured through the 
Shannon’s entropy, while \citet{barchi2020machine} introduced $\texttt{G}_\texttt{2}$, a parameter grounded in the Gradient Pattern Analysis method \citep{ROSA2008844, rosa2018gradient}. This latter metric relates to the asymmetric vectorial field denoting flux variation throughout the image. Notably, these supplementary metrics have exhibited greater efficacy in distinguishing ellipticals from spirals compared to the original \texttt{CAS} parameters \citep{rosa2018gradient}. When employed individually or in combination, these metrics offer an encompassing perspective on a galaxy's morphology. Further enhancing their utility, these metrics can synergize with machine learning algorithms to automate the classification workflow \citep{barchi2020machine, tohill2021quantifying}.

\begin{figure}
	\includegraphics[width=\columnwidth]{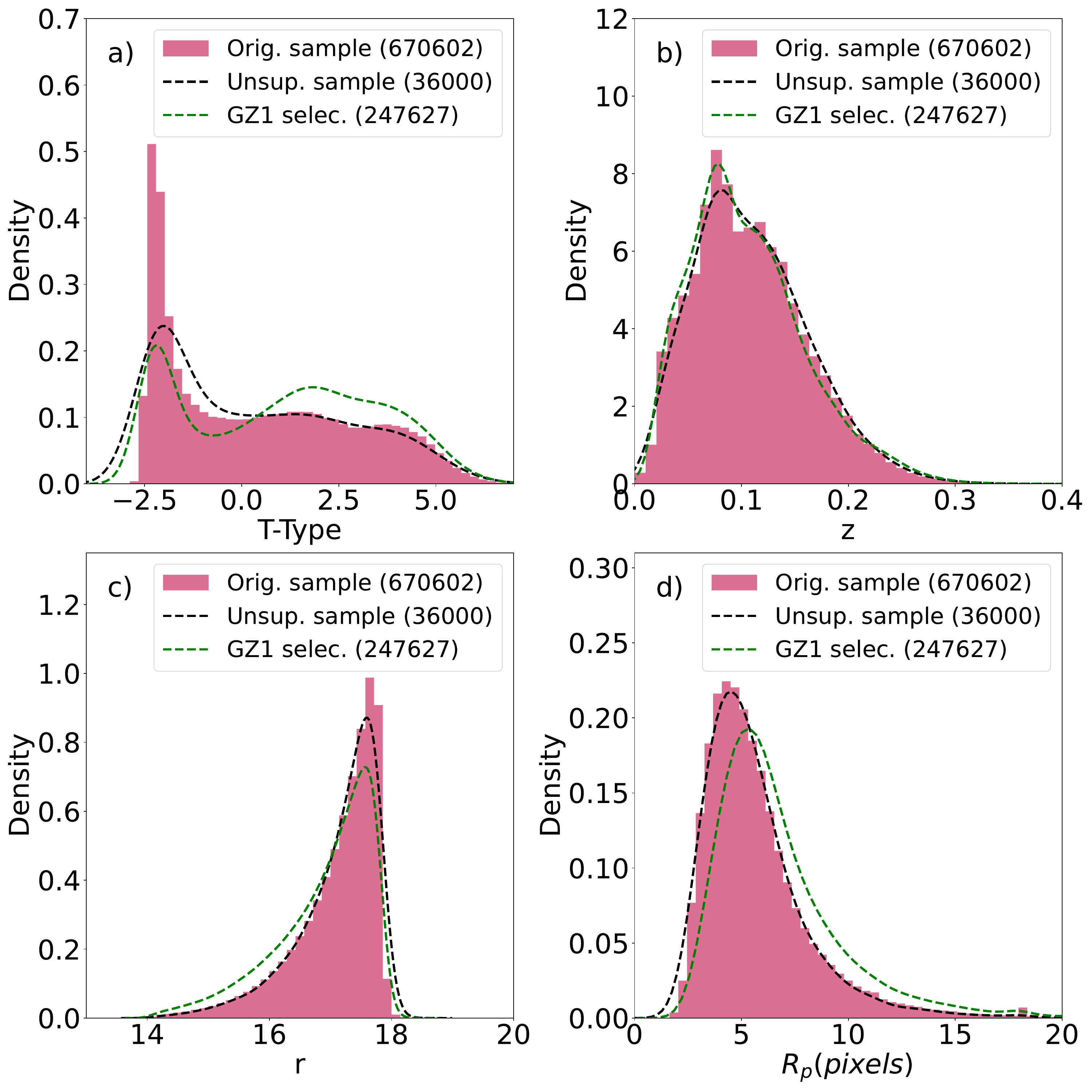}
    \caption{Distributions of the \texttt{T-Type} (a), redshift (b), magnitude (c) and petrosian radius (in pixels) (d). On each panel, we show original distribution from \citet{dominguez2018improving} in pink. PDF lines showcase our data selection. Black one represents the data we selected from the catalog for unsupervised clustering. All four panels show the match of the randomly selected data to the original dataset. Green PDF line corresponds to data selected from \citet{lintott2011galaxy}. Notable, selection from GZ1 contains slightly bigger, brighter and more \texttt{T-Type} contrasting galaxies.}
    \label{fig:data_combined}
\end{figure}

In our methodology, we harness the capabilities of the self-organizing maps (SOM) clustering algorithm \citep{kohonen1990self}, as implemented in the SOMbrero package \citep{villa2017stochastic}. Leveraging large-scale imaging surveys and multi-dimensional datasets, we move beyond traditional visual classifications to explore the morphological spectrum. The implications of employing unsupervised methods in galaxy morphology go beyond taxonomic revisions. This approach has the potential to unveil previously undiscovered evolutionary trajectories and illuminate the underlying physical processes that drive galaxy formation and transformation. Moreover, it lays the foundation for comparative analyses across diverse populations, offering insights into environmental impacts and cosmic assembly histories. Alongside clustering, we utilize robust Convolutional Neural Networks (CNNs) reinforced by transfer learning, label smoothing, and augmentation techniques to streamline and scale up the classification of larger data samples. Our proposed methodology unfolds in a horizontal manner, with each successive layer building upon its predecessor. The pipeline commences with image acquisition, segmentation, and metric extraction. Following this, we harness unsupervised techniques to discern the two most prominent morphological types in the dataset, relying solely on the extracted morphological metrics. These discerned types subsequently serve as the foundation for training a supervised classifier. Our hybrid framework aspires to capitalize on the merits of both methodologies: utilizing unsupervised analysis to illuminate inherent structures, and harnessing supervised learning to refine classification precision with the potential to scale to even more extensive datasets. Through the amalgamation of these techniques, we aim to bolster the accuracy and efficiency of galaxy morphology classification in an era increasingly steered by data-driven insights.

This paper is structured as follows: In Section \ref{sec:data} we present our datasets and describe the rationale motivating used data selection. In Section \ref{morpology_comparison} we compare morphology classification systems, detailing the reasons for methodology choice. In Section \ref{method} we present, in detail, the approach of unsupervised morphological clusterization and hybrid classification of galaxies.  In Section \ref{section_sdss_hst} we provide an extension of this method applied on the HST data, comparing it with SDSS. Section \ref{section_conclusions} draws conclusions on the presented method, its application, and obtained results.
\section {Sample Selection and Data Used}\label{sec:data}

In this research, we utilize several data samples from SDSS Data Release 7 \citep{abazajian2009seventh} for training and testing our approach. The samples are composed of images of galaxies in the r-band, with a redshift range of $0.01 < z < 0.3$, Petrosian magnitude in the r-band brighter than 17.78 (spectroscopic magnitude limit), and $|b| \geqslant 30^\circ$, where b is the galactic latitude. Our primary data sample is the dataset used in the \citet{dominguez2018improving} paper. This catalog contains 670,602 galaxies, all of which have value of \texttt{T-Type}, that 
is to
be used to define its morphology. 
Using this catalog
we randomly select 5\% (36,000, maximum amount for reasonable processing time) of the data without replacement. This step results in the creation of two datasets: (1) one used for an unsupervised clustering in order to extract morphological metrics and, based on these metrics, generate morphological labels via an unsupervised algorithm, and (2) one for supervised testing, to be used for the testing of the method through supervised classification. This selection process is shown in Figure \ref{fig:data_combined}. The unsupervised dataset is used to train a supervised Convolutional Neural Network (CNN). In the practical application on raw data, this step would not involve true labels. However, for the purposes of this research, we use them to compare and validate that the proposed method is functioning as expected. The supervised or testing dataset is reserved for the evaluation of the trained model. Importantly, these two datasets do not overlap, but both encompass the same domain of redshift and magnitude.

In addition to our primary dataset, we also incorporate 
labels assigned by the Galaxy Zoo 1 (GZ1) project \citep{lintott2011galaxy}
. We select galaxies that were marked as confident and exclude those used for supervised training, resulting in a separate GZ1 dataset for testing the model. This dataset is to be used as an additional test metric in an attempt to compare method performance on \texttt{T-Type} and human-based classifications. The selection is shown in Figure \ref{fig:data_combined}.
The intersection between \citet{dominguez2018improving} and GZ1 lies not just in the galaxies they study, but also in the distinct methodologies they employ. GZ1 stands as a pioneering endeavor, where visual classifications from volunteers were harnessed to curate a comprehensive catalog. This catalog delineated galaxies based on their morphology as either Spirals or Ellipticals and additionally annotated the confidence level associated with each classification. The volunteer-driven approach ensured a robust and diverse dataset, as each galaxy underwent manual scrutiny. On the other hand, \citet{dominguez2018improving} built upon the GZ1 catalog, utilizing it as a foundation for their supervised machine learning algorithm. They aimed to extend classifications to galaxies that had been ambiguously categorized in the GZ1 catalog and to introduce the \texttt{T-Type} metric. This metric, a numerical scale, provides a more detailed morphological characterization, augmenting the traditional classification system. Consequently, we have two datasets that we aim to replicate, one based on T-Type and another on visual classification.

Additionally, as our pipeline employs two different approaches: (1) unsupervised Machine Learning based on morphology and (2) supervised Deep Learning, the types of input data vary accordingly. For (1), we utilize Flexible Image Transport System (\texttt{.fits}) images and a detailed preprocessing pipeline. Using \texttt{.fits} images ensures that the data remains in its most raw and comprehensive form, allowing the unsupervised algorithm to capture the nuances and subtleties of galaxy morphologies. For the deep learning phase, our data choice diverges significantly. Default input for the CNN model consists in common image formats (such as \texttt{.png}, \texttt{.jpg} or \texttt{.tiff}), so instead of the more detailed \texttt{.fits} images, we opt for \texttt{.jpg} images sourced directly from the SDSS (Sloan Digital Sky Survey) website. We download images with resolution 448px$\times$448px, and scale calculated by: \( scale = R_p \times 0.02\), where $R_p$ is Petrosian radius in arc-seconds. This combination provides a balance between size of the image and amount of details around the galaxy. Our approach to preprocessing these images is characterized by its minimalism. We eschew extensive manipulations, relying on the power of deep learning to automatically extract and learn important features from the images.
\section{Morphological system comparison - \texttt{CAS} vs  \texttt{EGG}} \label{morpology_comparison}

\begin{figure}
	\includegraphics[width=\columnwidth]{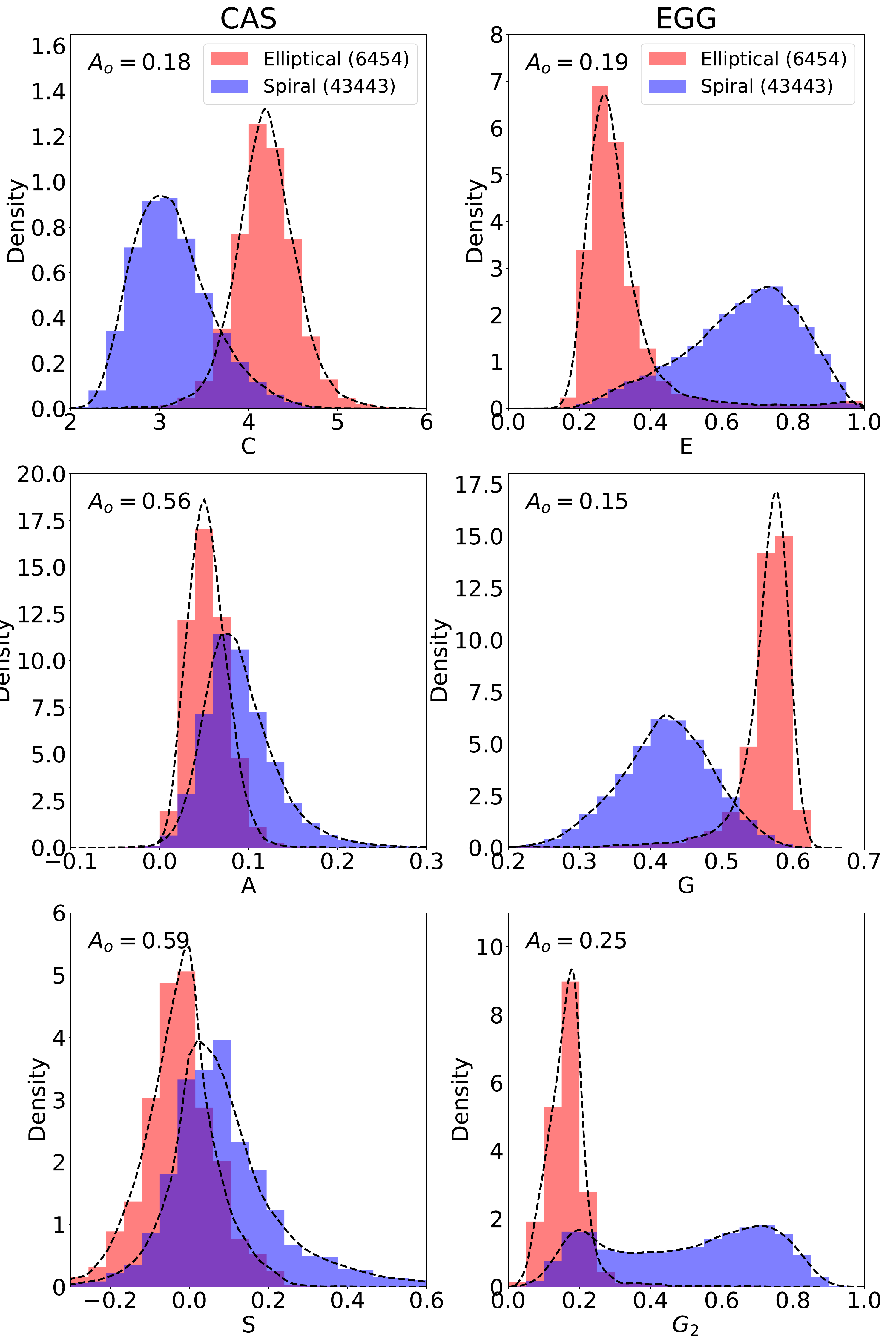}
    \caption{Visual comparison of \texttt{CAS}  and \texttt{EGG} systems based on respective metric values when using curated data sample of 49,897 galaxies. We plot the distribution of values of each metric for a given class (i.e. Elliptical or Spiral) and visually inspect how well this distributions are separated. $A_o$ provides the value of overlap between distributions that are normalized by area. }
    \label{fig:cas_vs_egg}
\end{figure}

The study of galaxies has long been enriched by various classification systems, aiming to understand their diverse structures and evolutionary histories. However, the search for the most meaningful and reliable set of parameters reflecting the morphological context of a galaxy is still ongoing. In this section, we present a comprehensive comparison between two morphological classification systems, focusing mainly on how their metrics could effectively distinguish between early and late-type galaxies using our imaging. In the following, we briefly describe the well-known morphological system, \texttt{CAS} (Concentration, Asymmetry, and Smoothness) \citep{conselice2000asymmetry, conselice2003relationship}; and the new one we propose here, \texttt{EGG} (Entropy, Gini, and Gradient Pattern Analysis (GPA), second moment).

\begin{figure}
	\includegraphics[width=\columnwidth]{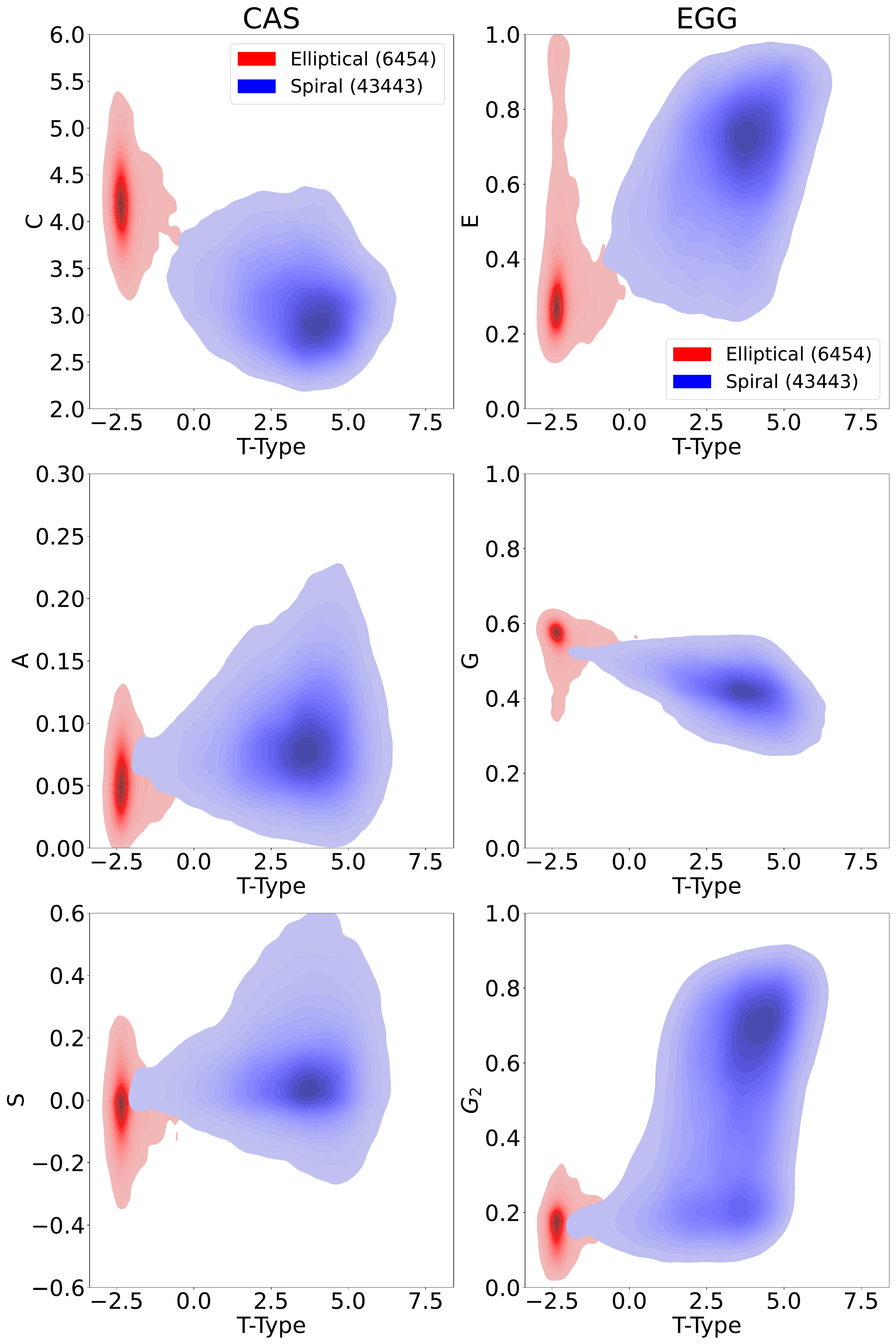}
    \caption{Visual comparison of \texttt{CAS} and \texttt{EGG} systems based on respective metric values versus \texttt{T-Type}  when using curated data sample of 49,897 galaxies.}
    \label{fig:cas_vs_egg_ttype}
\end{figure}

\subsection{\texttt{CAS}}
This system quantifies three fundamental aspects of a galaxy's stellar light distribution: Concentration (\texttt{C}), Asymmetry (\texttt{A}), and Clumpiness (\texttt{S}) \citep{conselice2000asymmetry,conselice2003relationship}. We briefly describe them, letting the readers find specific details in the vast literature on the topic:

\begin{itemize}
\item \textbf{Concentration (\texttt{C})}: The concentration of a galaxy's light quantifies the centralized distribution of stellar light within a galaxy. Mathematically, the concentration index is defined as:
\[
C = 5 \times \log \left(\frac{R_{80}}{R_{20}} \right),
\]
where \( R_{80} \) and \( R_{20} \) are the radii containing 80\% and 20\% of a galaxy's total light, respectively, within 1.5 times the Petrosian radius \citep{kent1985analysis, bershady2000structural}. The literature contains different approaches to calculate concentration. Here we follow the method proposed in \citet{conselice2003relationship} and \citet{lotz2004new}. Additionally, several studies used different pairs of $R_1$ and $R_2$ \citep{lotz2004new, ferrari2015morfometryka}. In our case, we use default values for outer and inner radii with $R_1 = 80 \%$ and $R_2 = 20 \%$ proposed in \citet{conselice2003relationship}.

\item \textbf{Asymmetry (\texttt{A})}: The asymmetry of a galaxy is a measure that quantifies its deviation from bilateral symmetry. This is computed by rotating the galaxy image by \(180^\circ\), subtracting the rotated image from the original, and then normalizing the result to the sum of the pixel intensities in the original galaxy image. Mathematically, 
\[
A(\phi) = \frac{\sum |I_{\text{original}} - I_{\text{rotated}}|}{\sum I_{\text{original}}} - A_{bkg},
\]
where \( I \) represents the pixel intensity, and $A_{bkg}$ is asymmetry of background. A higher value of \texttt{A} indicates a greater degree of deviation from symmetry. For example, spiral galaxies, which often feature arms and bars, generally exhibit higher asymmetry compared to elliptical galaxies.

\item \textbf{Clumpiness (\texttt{S})}: The clumpiness (sometimes also called smoothness in literature, \citep{barchi2020machine}), of a galaxy's light distribution characterizes its granularity or patchiness, especially at high spatial frequencies.  To quantify this, the clumpiness index \texttt{S} is defined as the ratio of the amount of light contained in high-frequency structures to the total light of the galaxy. Computationally, \texttt{S} is determined by :
\[
S = 10 \times \left( \frac{\sum |I_{\text{original}} - I_{\text{smoothed}}|}{\sum I_{\text{original}}} \right) - S_{bkg},
\]
where \( I_{\text{original}} \) is the pixel intensity of the original galaxy image and \( I_{\text{smoothed}} \) is its smoothed version, and $S_{bkg}$ is smoothness of background. A higher \texttt{S} value indicates a more clumpy or patchy light distribution \citep{conselice2003relationship}.

\end{itemize}

\subsection{\texttt{EGG}}

Here, we introduce an alternative approach that, through testing, demonstrated improved separation between two morphological classes across all three metrics compared to the \texttt{CAS} system, on SDSS data, as shown in this section. We term this system "EGG," encompassing the metrics of Entropy, Gini, and Gradient Pattern Analysis (second moment). While each metric has been used individually or as part of various systems, we combine them here. Our motivation primarily stems from their collective ability to distinguish effectively between the two major morphological classes: ETGs and LTGs.

\begin{itemize}
    \item \textbf{Entropy (\texttt{E})}: The entropy of information, represented as \texttt{E}, characterizes the distribution of pixel values in an image. Originating from Shannon entropy, it captures the randomness or unpredictability inherent in the image's information content. In essence, it mirrors the concentration by assessing pixel density/frequency across specified bins. The number of entropy bins can be adjusted to better fit the data, determining how the flux distribution is divided. 
    Mathematically, entropy \(E(I) \) for a random variable \( I \) is:
    \[
        E(I)=-\sum_{k}^{K}\,p(I_{k})\,\log[p(I_{k})],
    \]
    with \( p(I_k) \) as the likelihood of the occurrence of value \( I_k \), \( k \) denoting a specific value, and \( K \) being the total bin count. Entropy achieves its peak with a uniform distribution and its lowest with a delta function. The normalized entropy is given by:
    \[
        \tilde{E}(I) = \frac{E(I)}{E_{\text{max}}},
    \]
    where \( 0 \leq \tilde{E}(I) \leq 1 \) and \( E_{\text{max}} = \log(K) \).
    Elliptical galaxies, due to their naturally smooth flux distribution, typically exhibit lower entropy. In contrast, spiral galaxies, known for their irregular structures and inherent pixel value heterogeneity, tend to display higher entropy values \citep{biship2007pattern, ferrari2015morfometryka}.

    \item \textbf{Gini Coefficient (\texttt{G})}: Originally used in economics to represent wealth distribution, the \texttt{G} coefficient has been adapted for galaxy morphology to measure the relative flux distribution across pixels corresponding to a galaxy. While it correlates with concentration, it does not necessarily presume the brightest pixels to be centrally positioned in the galaxy image \citep{abraham2003}. \texttt{G} is defined mathematically for a discrete population as: 
    \[
        G = \frac{1}{2 \bar{X} n(n - 1)} \sum_{i=1}^{n} \sum_{j=1}^{n} |X_i - X_j|,
    \]
    with \( n \) denoting the galaxy's pixel count and \( X_i \) signifying the flux value of the \( i^{th} \) pixel. An efficient computation, after sorting \( X_i \) in an ascending sequence leads to:
    \[
        G = \frac{1}{\bar{X} n(n - 1)} \sum_{i=1}^{n} (2i - n - 1)X_i.
    \]
    \texttt{G} correlates with the concentration index, 
    increasing
    with the increasing light fraction in a compact component \citep{abraham2003}. Elevated \texttt{G} values can occur when the galaxy's brightest pixels are not centrally situated, differentiating it from concentration \citep{lotz2004new}. A high \texttt{G} implies that the galaxy's overall light predominantly resides in few pixels—often characteristic of the Elliptical class. Conversely, a low \texttt{G} suggests an even light distribution across pixels, aligning more with the Spiral class. This provides a measurable metric to assess a galaxy's light concentration and distribution, offering insights into its structural, morphological, and formative history.

    \item \textbf{Second Gradient Moment (\(\texttt{G}_\texttt{2}\)}): As \citet{teseRubens} points out, the Gradient Pattern Analysis (GPA) comprises four moments, but only the first and second are relevant for galaxy morphology classification. \citet{rosa2018gradient} demonstrated that an enhanced and updated version of the second moment (with the necessary adjustments) surpasses the traditional \texttt{CAS} classification for galaxy separations \citep{conselice2003relationship}. 
    Essentially, GPA is a method crafted to gauge gradient bilateral asymmetries in a numerical grid. Within galaxy morphometry, the second gradient moment, represented as \(\texttt{G}_\texttt{2}\), stands out. It proves especially valuable, showcasing its potential to differentiate ETGs from LTGs more effectively than traditional morphometric metrics. For example, in an analysis leveraging the SDSS-DR7 dataset, \(\texttt{G}_\texttt{2}\) registered an estimated separation efficiency of approximately 90\% between galaxy categories \citep{rosa2018gradient}.

    Primarily, the extraction process of \(\texttt{G}_\texttt{2}\) entails identifying pairs of pixels at an equal distance from the center and comparing their modulus (strength) and phase (direction). Symmetric pixel pairs, those with identical modulus but opposing phases, are recognized and discarded. This operation is carried out for all unique pixel pairs. The resulting matrix is termed an asymmetry vector field, given that all symmetric pairs have been eliminated. Subsequently, we compute the count of asymmetric vectors, their summation, and the sum of their moduli. The $confluence$ is calculated 
    to determine if the vectors align and possess equivalent magnitude:

    \[
    confluence = \left(\frac{\left|\sum_{i}^{V_{A}} v_{a}^{i}\right|}{\sum_{i}^{V_{A}}\left|v_{a}^{i}\right|}\right),
    \]
    where, \(v_{a}\) represents the list of asymmetrical vectors, \(V_{A}\) denotes the count of asymmetric vectors.
    
    The final value for \(\texttt{G}_\texttt{2}\) is ascertained using \citep{rosa1999characterization, ramos2000generalized, rosa2003gradient, teseRubens}:

    \[
    G_{2} = \frac{V_{A}} {V} (1-confluence),
    \]
    where, \(V\) signifies the total number of pixels, \(V_{A}\) represents the count of asymmetric pixels, 1 acts as the normalization factor \footnote{Originally, in \citet{rosa2018gradient}, the value 2 was used instead.}.

\end{itemize}

Despite originally defined to a range [0,2], we renormalize it to the range [0,1]
This adjustment preserves the metric's efficacy while standardizing its range, thereby facilitating comparative evaluation alongside other \texttt{EGG} metrics.

\begin{figure}
\begin{center}
	\includegraphics[width=150px]{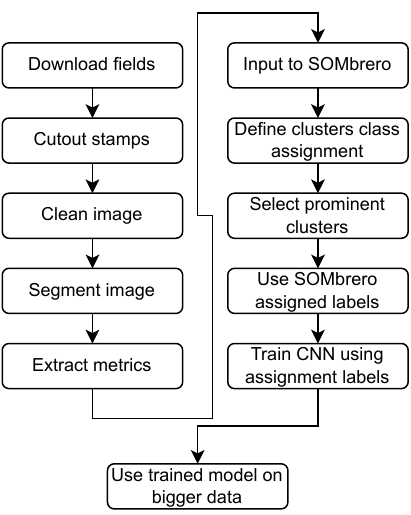}
    \caption{Diagram depicting steps of the proposed method we test in this paper.}
    \label{fig:diagram}
\end{center}
\end{figure}

For the morphological systems comparison, we use curated sample of galaxies featuring well resolved and bright subjects specifically selected from \citet{dominguez2018improving}. This galaxies were selected based on parameter \texttt{K} as the area of the galaxy’s Petrosian ellipse divided by the area of the Full Width at Half Maximum (FWHM), defined as:

\[
    K =  \left(\frac{R_p}{FWHM/2}\right),
\]
where $R_p$ is the Petrosian radius (see \citet{petrosian1976surface}; \citet{eisenstein2011}, for
more details about $R_p$). Our testing sample restrict value of $K \geq 20$ \citep{barchi2020machine}.
Figures \ref{fig:cas_vs_egg} and \ref{fig:cas_vs_egg_ttype} visually contrast the two systems using a selection of 49,897 galaxies, showcasing distinctive morphological features. In Figure \ref{fig:cas_vs_egg}, the capacity of each metric to segregate known morphological classes is evident, with \texttt{EGG} offering more pronounced differentiation. Figure \ref{fig:cas_vs_egg_ttype} extends this analysis, incorporating the \texttt{T-Type} dimension to visualize metric-based galaxy separations.

To enhance our visual analysis with quantitative insights, we compute the overlap area of the distributions and define it as :
\[
A_o = \sum_{i=0}^{N-1} \min\left(\text{hist}_1(i), \text{hist}_2(i)\right) \times \left(\text{bin\_edges}(i+1) - \text{bin\_edges}(i)\right),
\]
where \(N\) is the total number of bins, defined by Freedman-Diaconis Rule \citep{freedman1981histogram}, \(\text{hist}_1(i)\) and \(\text{hist}_2(i)\) represent the heights of the \(i\)th bin for \(data_1\) and \(data_2\), respectively, \(\text{bin\_edges}(i)\) represents the edges of the bins.

Given that both distributions in each plot are area-normalized, this overlap measurement offers a consistent comparative index across metrics. While we recognize its inherent limitations, such as the varied metric ranges precluding a direct comparison of peak distances, \( A_o \) in Figure \ref{fig:cas_vs_egg} objectively quantifies the separation capabilities of each metric.

\begin{figure}
	\includegraphics[width=\columnwidth]{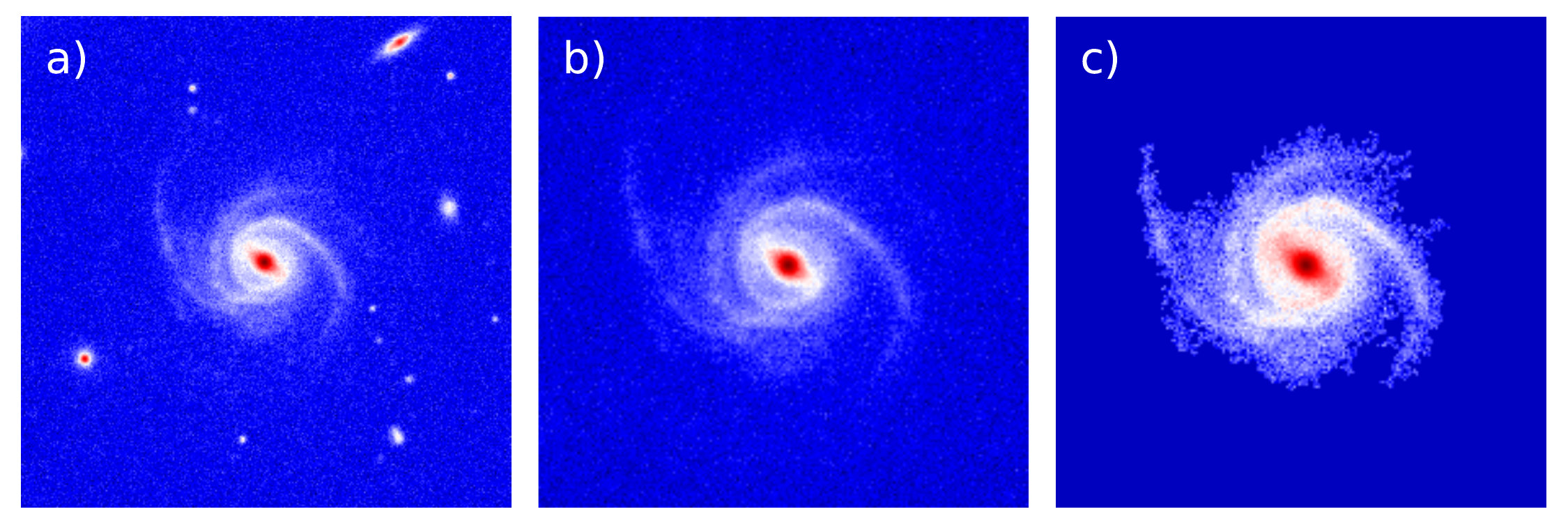}
    \caption{Processing Pipeline for Metric Extraction in Unsupervised Machine Learning: Panel (a) displays the original cutout of the galaxy. Panel (b) presents the cutout with secondary sources removed, and Panel (c) shows the segmented cutout, which is the version used for the \texttt{EGG} system. Important to note that on Panel (c) all pixels that do not belong to the galaxy, assigned as 0.}
    \label{fig:segmentation_example}
\end{figure}

An interesting observation pertains to the bimodality of the \(\texttt{G}_\texttt{2}\) distribution, particularly the peak around \(\texttt{G}_\texttt{2} = 0.2\) and $\texttt{T-Type} = 0$ in Figure \ref{fig:cas_vs_egg}. Crossmatching this peak with \texttt{T-Type} vs \(\texttt{G}_\texttt{2}\)) in Figure \ref{fig:cas_vs_egg_ttype} reveals that this galaxy group is consistent with the lenticular type. This observation suggests \(\texttt{G}_\texttt{2}\)'s refined capability in detecting subtle differences in gradient fields and discerning LTG subtypes.

We intend to exploit this segregation by inputting the metric values into an unsupervised clustering algorithm. This strategy aims to unearth intrinsic metric interrelations, ultimately enabling unsupervised galaxy classification based on morphological metrics alone. The evidence presented underscores the \texttt{EGG} system's potential as a promising tool in galaxy morphological classification. 
Given its efficacy, we utilize \texttt{EGG} as our cornerstone analytical tool for this paper. We base our hybrid galaxy classification approach on this system and, to gauge its versatility, we apply this method to data from diverse astronomical surveys. Through these endeavors, we aim to solidify \texttt{EGG}'s broad-spectrum applicability. 
\begin{figure}
    \begin{center}
	\includegraphics[width=220px]{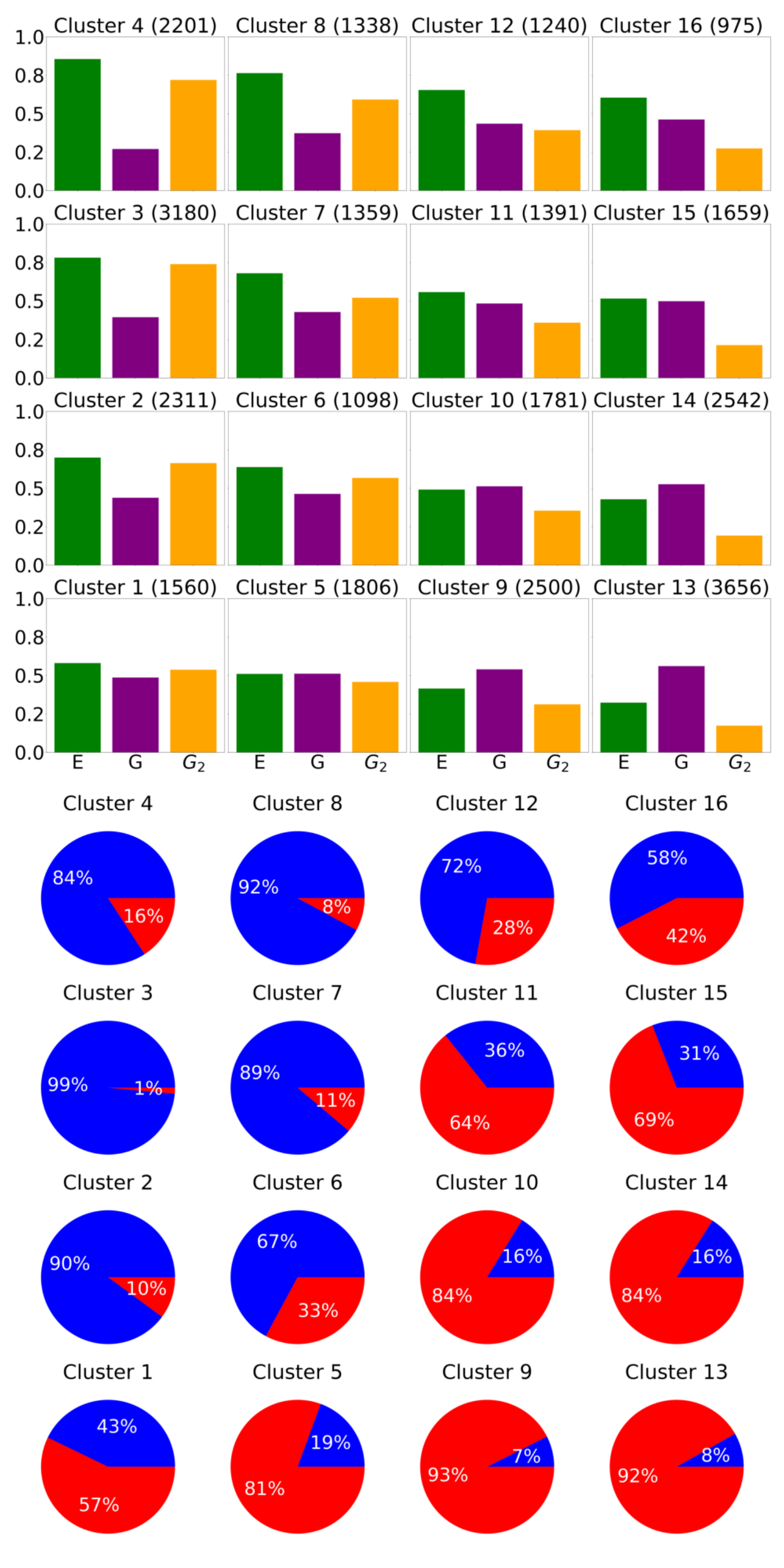}
    \end{center}
    \caption{Top panel: This panel showcases the grid resulting from inputting the tabular metrics data into the SOMbrero algorithm. The title displays the cluster label and the number of subjects within that cluster. Each bar signifies the mean metric value within the respective cluster. Bottom panel (red: elliptical; blue: spiral; labels used from original catalog): This section incorporates a label variable to inspect the algorithm's galaxy distribution, considering each galaxy's true class. The pie chart indicates the strong correlation between the metrics and morphological classes. The distribution is based on metric values, and each cluster hosts varying galaxy counts from the total 36,000.}
    \label{fig:final_plot_before_sc}
\end{figure}

\section{Unsupervised morphological classification}\label{method}
In this paper, we propose a method capable of rapidly generating a binary-labeled catalog for Spiral and Elliptical classes based on non-parametric morphological parameters extracted from galaxy images.
We employ an unsupervised machine learning (UML) algorithm to harness the information embedded within 
\texttt{EGG} system.
This choice was based on the reported strong correlation between the metrics and specific physical properties (see Section \ref{morpology_comparison}). Therefore, using these metrics as input for an unsupervised clustering algorithm not only yields high accuracy in classification but also provides a robust foundation grounded in tangible, well-defined metrics, each corresponding to a distinct characteristic of a galaxy.

In this section, we offer a thorough overview of our unsupervised methodology, detailing the algorithms used, the preprocessing steps undertaken, and the rationale behind our choices. We emphasize the benefits of this data-centric strategy while recognizing the challenges and constraints inherent in such a paradigm. By leveraging the capabilities of unsupervised methods, we aim to make a contribution to the landscape of galaxy morphology analysis.

\subsection{Hybrid Approach in a Nutshell}
In the testing phase of our methodology, we aim to reproduce both \texttt{T-Type} and visual classification catalogs. Simultaneously, we seek to compare the efficiency of unsupervised classification techniques with the results obtained from traditional machine learning and deep learning approaches, as documented in existing literature. We focus our primary data sample and evaluation on the SDSS dataset from \citet{dominguez2018improving}, without imposing limits on redshift or magnitude. This strategy highlights the resilience of our method, emphasizing its potential to handle more intricate and subtle datasets.

To encapsulate our method, the process involves: (1) data preprocessing, which encompasses data download, galaxy cutout, cleaning, and segmentation; (2) extraction of metrics from the preprocessed stamps; (3) application of UML to establish a clustering hierarchy; (4) analysis of this hierarchy to allocate labels to subjects based on their classes; and (5) selection based on morphological prominence. Finally, in step (6), we train a CNN model in a supervised manner, using the classifications (labels) generated by the unsupervised algorithm. We then test this model on larger catalogs to assess its performance and wider applicability in real-world scenarios. Below, we delve deeper into each of these stages. These steps are illustrated in Figure \ref{fig:diagram}.

\subsection{Unsupervised Machine Learning} 
UML is an approach in which the algorithm is tasked with deciphering patterns in data without prior knowledge of the ground truth. Typically known as clustering, this process involves the algorithm segmenting the input data into an arbitrary number of clusters based on inherent relationships within the data. This approach stands in stark contrast to Supervised Machine Learning (SML), where the model is provided with the ground truth and refines its weights by comparing its predictions against it. The effectiveness of the clustering is deeply dependent on the quality of the input data.

Numerous unsupervised algorithms are available for clustering, including K-Means \citep{likas2003global}, OPTICS \citep{ankerst1999optics}, and Autoclass \citep{cheeseman1988autoclass}, to name a few. In this study, we chose SOMbrero \citep{villa2017stochastic} as our primary clustering algorithm. Main motivation for this choice being robust visualization and analysis tools available. Furthermore, we present several alternative methods to provide comparative results, eliminating any algorithmic dependency, more precisely Agglomerative Hierarchical Clustering \citep{day1984efficient}, Gaussian Mixture Clustering \citep{ouyang2004gaussian} and Spectral Clustering \citep{von2007tutorial}. For three algorithms, we used the implementation from \citet{scikit-learn}. These alternative algorithms are detailed in Appendix \ref{sec:appendix1}.

\subsubsection{SOMbrero}
Self-Organizing Maps (SOM) \citep{kohonen1990self, kohonen1982analysis, kohonen1982self} are algorithms broadly utilized for clustering and visualization tasks. There are numerous implementations of the SOM algorithm across different mathematical and statistical platforms \citep{vettigliminisom, kohonen2014matlab}. SOMbrero is an R package \citep{villa2017stochastic} that offers a stochastic version of the SOM algorithm. This type of artificial neural network is trained using unsupervised learning to produce a low-dimensional, discretized representation of the input space of training samples, termed a map. The stochastic, or "on-line", version of the SOM algorithm is especially apt for managing large datasets since it updates the map iteratively for each data point instead of batch processing the complete dataset. The primary strength of SOMbrero lies in its ability to process numeric datasets. It achieves this through a suite of functions tailored for standard SOM operations. These functions enhance the efficient analysis and visualization of high-dimensional data, revealing intricate patterns and structures within the dataset.

\subsection{Data processing} \label{data_processing}
The method we propose for unsupervised morphological classification, rooted in morphological metrics, requires a systematic processing pipeline to extract metrics from the input images. 

The process begins with the download of a SDSS image of the field from the survey database. The subsequent step is the creation of a specific-sized cutout, designed to ensure precise background estimation. For example, an overly small cutout might lead to inaccurate background estimation, while an excessively large one could result in prolonged processing times. The cutout's size, in pixels, is given by: 
\[ cutout\_size (px) = scale \times r_{petrosian}, \]
where we use \( scale=10 \) as a constant multiplier to strike a balance to preserve reasonable amount of sky in the cutout, without adding excessive amount processing time. Galaxies that does not fit this criteria are discarded.

The subsequent pipeline stage encompasses the task of cleansing the secondary objects within the cutout, a necessary procedure to mitigate potential biases in the estimated parameters. This endeavor entails the replacement of pixels associated with secondary objects with Gaussian random values that correspond to the isophote of the primary object at the specified radial distances. By doing so, the influence of proximate secondary objects upon the primary object is minimized. After the cleaning, the third step is the production of a segmentation image, effectively delineating the pixels that are attributed to the galaxy. The primary challenge at hand is to devise a quantitative criterion for ascertaining which pixels encompass the luminosity emanating from the galaxy, in a manner that remains invariant to the redshift of the objects in consideration. One viable approach to tackle this challenge is to adopt the methodology adopted in \citet{lotz2004new}. Succinctly, this methodology entails the identification of galaxy pixels as those pixels exhibiting intensities surpassing a threshold defined as the intensity at the Petrosian radius (denoted as $\rm I(R_{p})$).

In summary, Figure \ref{fig:segmentation_example} delineates the processing pipeline. Panel (a) showcases the original cutout of the galaxy, while Panel (b) displays the cleaned stamps, which are images devoid of secondary sources. Panel (c) introduces samples of images purposed for morphological metric extraction, which subsequently feed into the unsupervised machine learning (UML) algorithm. 

\begin{figure}
    \begin{center}
	\includegraphics[width=220px]{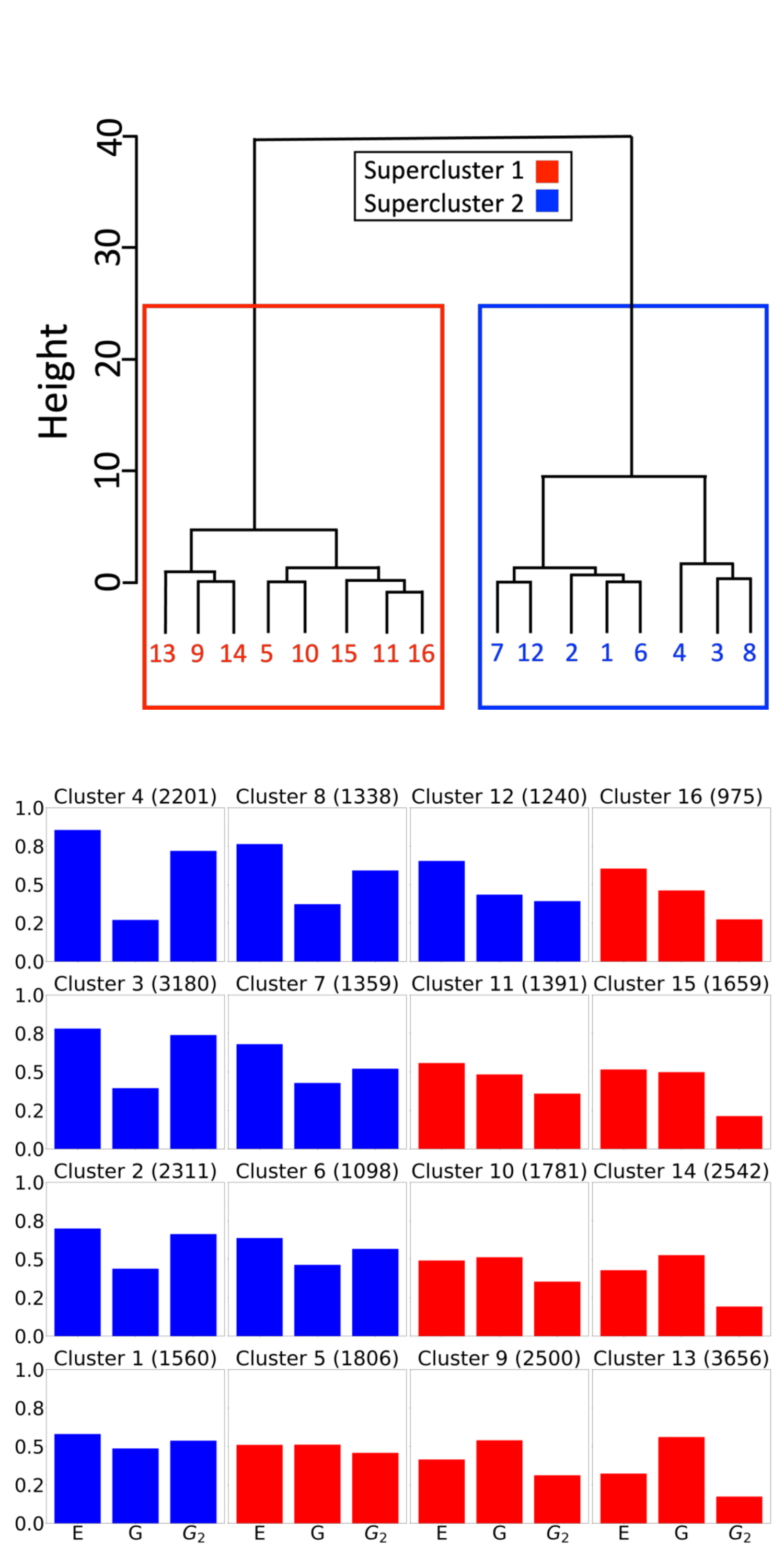}
    \end{center}
    \caption{Top panel: The dendrogram depicted here showcases the hierarchy of clusters and offers insights into the unsupervised algorithm's interpretation of the data. This plot allows us to deduce the number of superclusters we might want to use to devide the clusters. In this instance, we opted for two. Bottom panel (red: elliptical; blue: spiral; labels used from supercluster hierarchy): This figure mirrors the one shown in Figure \ref{fig:final_plot_before_sc} but now integrates the supercluster partitioning derived from the top panel.}
    \label{fig:final_plot_after_sc}
\end{figure}

\begin{figure*}
	\includegraphics[width=505px]{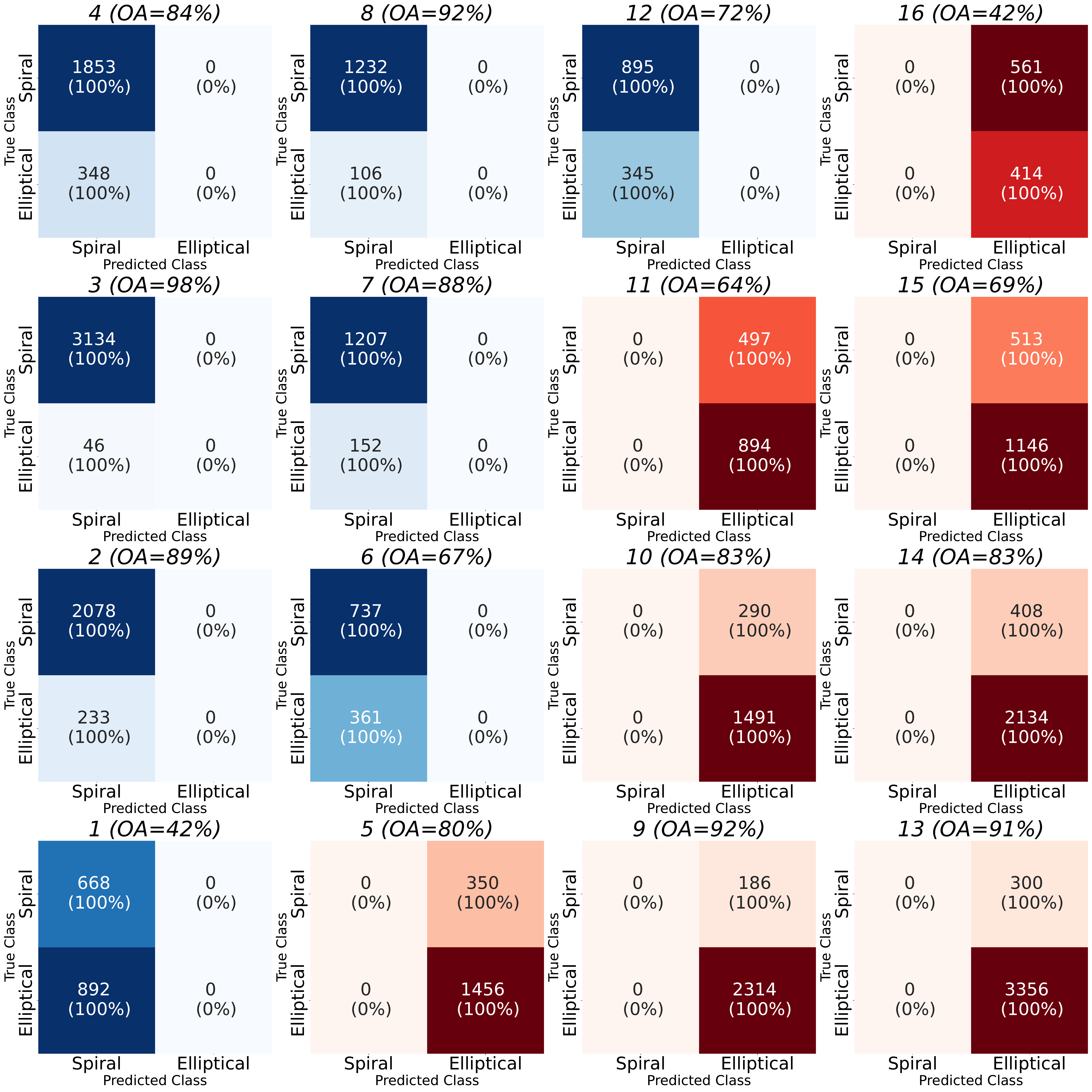}
    \caption{A grid of confusion matrices (CM) portraying the performance of the unsupervised classification for each distinct cluster. Title of each CM contains cluster id and Overall Accuracy inside this cluster. Colors are derived from supercluster hierarchy, and correspond to red - elliptical; blue - spiral. Both colors and the cluster distribution are preserved in relation to Figures \ref{fig:final_plot_before_sc} and \ref{fig:final_plot_after_sc}.}
    \label{fig:cm_mosaic_grid}
\end{figure*}

\subsection{Building SOMbrero cluster grid} 

After completing the segmentation process, we advance to the extraction of metrics. For each subject galaxy, we compute a suite of morphological metrics derived exclusively from the input image.
\begin{figure}
	\includegraphics[width=\columnwidth]{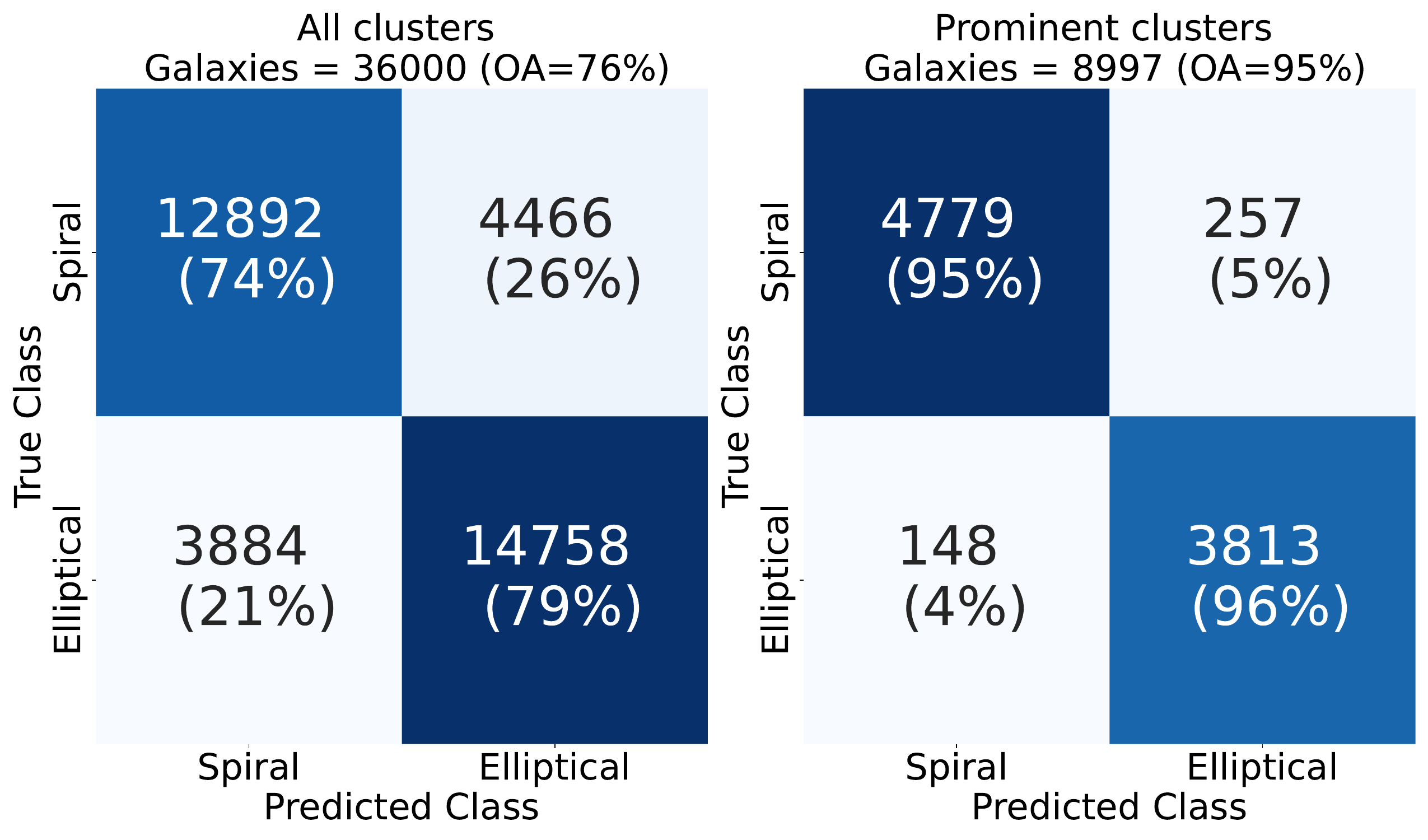}
    \caption{Comparison of the OA between the entire dataset's clustering and the prominent clusters. The two confusion matrices elucidate the disparity when considering the entirety of clusters from the SOMbrero grid versus only the prominent ones.}
    \label{fig:selection_proportions}
\end{figure}

\begin{figure*}
	\includegraphics[width=505px]{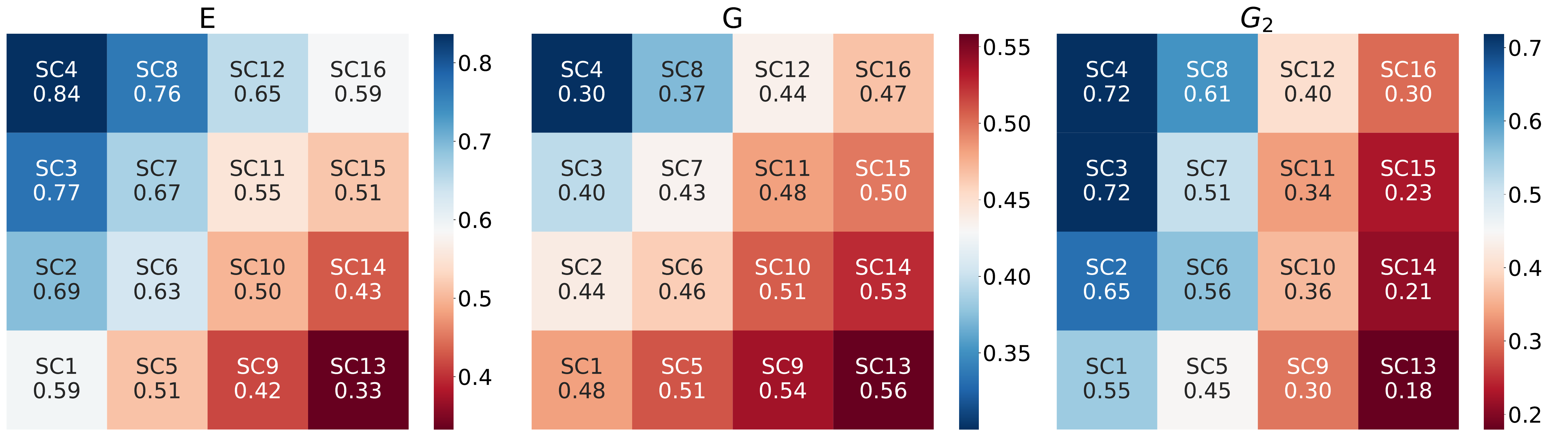}
    \caption{Mean metric values for each BMU (cluster). SOMbrero produces a matrix for each feature (or in our case, metric) that is input. Consequently, we discern three panels for \(\texttt{G}_\texttt{2}\), \texttt{G}, and \texttt{E}. Redder hues indicate a tilt towards the Elliptical class, while bluer hints at the Spiral class.}
    \label{fig:mean_by_cluster}
\end{figure*}

Upon metric extraction, we have morphological metrics presented in a tabular structure. In this structure, each row represents a unique galaxy, while each column corresponds to a particular metric. 
Independent of the number of columns, our chosen algorithm processes them into a 2D grid of a user-specified size. Subsequently, galaxies are grouped into clusters based on metric values. Figure \ref{fig:final_plot_before_sc} showcases this grid, where, for illustrative purposes, we employed 36,000 galaxies with a 4$\times$4 grid \footnote{This grid size was selected for visual demonstrations of the method and will be changed further in the chapter for optimal functioning of the SOMbrero}. The right panel of the same figure presents the distribution of galaxies based on their morphological classes. Crucially, the algorithm was blind to these classes; we overlaid them post-hoc for visualization. Remarkably, using only the input metrics, the algorithm could infer latent relationships between metrics and data, segmenting them into distinct clusters.

However, this distribution does not directly translate to classification. A dendrogram offers insights into how the unsupervised algorithm comprehends the data. Relying on this dendrogram, we can determine an optimal number of superclusters that we believe best represents the underlying data or specific goals. Initially, we have 16 clusters; post-separation, this consolidates into two superclusters, each embracing several clusters. The left panel of Figure \ref{fig:final_plot_after_sc} exemplifies such a dendrogram.

It is paramount to understand that we are operating with incomplete information regarding the data at this juncture; our sole guide is the dendrogram. Referring to our given example, there exists an apparent division between the two primary branches. This distinction becomes the foundation to delineate our superclusters: Red and Blue. Henceforth, Supercluster 1 (SC1) amalgamates clusters 13,9,14,5,10,15,11, and 16, while SC2 integrates the residual clusters 7,12,2,1,6,4,3,8.

The right panel of Figure \ref{fig:final_plot_after_sc} mirrors the mosaic from Figure \ref{fig:final_plot_before_sc}. However, bars now bear colors aligned with their respective superclusters.
Given that we are harnessing morphological metrics, we can exploit their ties to the galaxies' physical properties to surmise the morphological class each supercluster is most akin to.

\subsubsection{Unsupervised Classification}

Let us focus on cluster 3, a part of Supercluster 2 (SC2) for illustration. The metrics of this cluster strongly hint towards a majority of spiral galaxies. Spiral galaxies typically display elevated entropy (\texttt{E}) values because of their irregular light distribution, relative to elliptical galaxies. They also tend to possess elevated \(\texttt{G}_\texttt{2}\) values due to their perturbed gradient field. The \texttt{G} values for spiral galaxies are often lower, reflecting their less uniform light distribution compared to their elliptical counterparts. These observations lead us to deduce that cluster 3, and by implication SC2, is predominantly populated by spiral galaxies. In contrast, cluster 9, a member of Supercluster 1 (SC1), evidences a different set of metric features. Galaxies within this cluster commonly have reduced \texttt{E} and \(\texttt{G}_\texttt{2}\) values but elevated \texttt{G} values, signaling a preponderance of elliptical galaxies. This indicates that SC1 is majorly composed of elliptical galaxies. Clusters 3 and 9 are particularly instructive for in-depth examination. We name them "prominent" clusters, these clusters are positioned at the corners of the grid, showcasing galaxies that exhibit stark differences in their metrics.

This characteristic clustering of prominent galaxies at the grid's extremities is largely a consequence of the Self-Organized Maps (SOM) mechanism. During the training phase, every data point in the dataset is examined to determine the node whose weight vector aligns most closely with the data point, as defined by a specific distance measure. This node becomes the Best Matching Unit (BMU). Adjacent nodes to the BMU are subsequently pinpointed. Initially, this neighborhood has an expansive size, which diminishes progressively. The weight vectors of the BMU and its neighboring nodes are adapted to better resemble the data point. A learning rate, which starts elevated and diminishes over time, determines this adjustment magnitude. This iterative process, encompassing competition, cooperation, and adaptation, recurs for every data point, often over multiple cycles. Gradually, the nodes "self-organize" to depict various regions or clusters within the input space.

Our subsequent move is to evaluate the accuracy of our catalog. Although real-world circumstances often do not allow for such verification, this exercise is pivotal for juxtaposing our method's efficacy with the ground truth labels. Figure \ref{fig:cm_mosaic_grid} unveils the results of our unsupervised algorithm, which relies exclusively on morphological metrics. The figure elucidates classification accuracy across the entire grid. Each individual cell hosts a confusion matrix, indicating accuracy per cluster. Corner clusters, which we term prominent, house the dominant share of galaxies and manifest the elevated accuracy. As one transitions towards the figure's core, a decline in both galaxy count and accuracy becomes evident, underscoring the limitations of our metrics in these clusters.

\subsubsection{Prominent cluster selection} 

\begin{figure*}
	\includegraphics[width=505px]{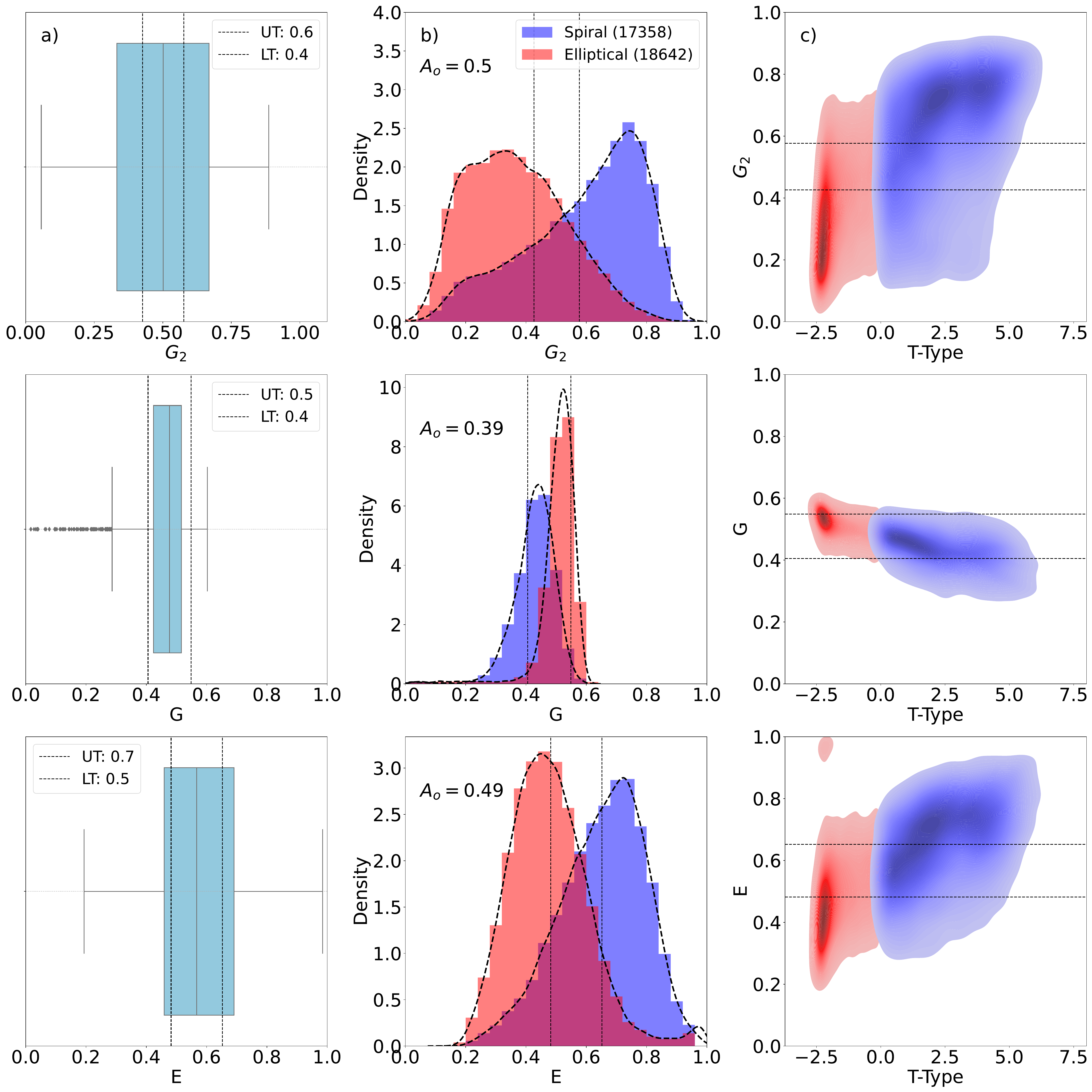}
    \caption{Visualization of prominent clusters based on three metrics. Each row corresponds to an input metric. The first row for \(\texttt{G}_\texttt{2}\) presents a boxplot of its values in Panel (a). Upper and Lower thresholds, UT and LT respectively, define prominence, with values outside these bounds considered prominent. Panels (b) and (c) for \(\texttt{G}_\texttt{2}\) overlay these thresholds on distributions by class and by \texttt{T-Type}. The thresholds capture the most distinct values both for the metric and \texttt{T-Type}, highlighting the pronounced differences between classes. After selection based on these thresholds, we preserve 5036 Spiral and 3961 Elliptical galaxies. $A_o$ provides the value of overlap between distributions that are normalized by area.}
    \label{fig:selection_metrics}
\end{figure*}

If we evaluate the Overall Accuracy (OA) and the misclassifications across the entire grid, we observe a relatively high number of instances of both classes being misclassified. The predictions provided by the SOMbrero algorithm indicate an important observation: while the algorithm was capable of discerning different galaxy types, it did not classify all instances correctly. In the left panel of Figure \ref{fig:selection_proportions}, we present the confusion matrix for the entire grid. These misclassifications are termed noise. Given this noise, using these values directly for supervised training could introduce complications in the Convolutional Neural Network (CNN). Such complexities might obscure its learning capability, hinting at potential for refinement.

To augment our results, increasing OA and reducing the noise, we retain only prominent clusters. Based on the morphological metrics, clusters 3, 4, and 8 mainly represent Spiral galaxies, while clusters 9, 13, and 14 are predominantly Elliptical. Visually pinpointing such clusters is relatively intuitive. However, ensuring scalability and adaptability to varied datasets demands an automated approach.

Our selection is rooted in galaxy metrics and their mean values per cluster (or BMU). As established, Spiral galaxies typically manifest high \(\texttt{G}_\texttt{2}\) and \texttt{E}, but low \texttt{G} values, with Elliptical galaxies displaying opposite trends. Leveraging this, we can selectively maintain clusters with these distinctive metric values. Figure \ref{fig:mean_by_cluster} showcases the mean metric distribution for each BMU (cluster). This distribution serves as our primary guideline for prominent cluster selection. The correlation between higher accuracy and distinct metric values becomes evident when juxtaposing Figures \ref{fig:mean_by_cluster} and \ref{fig:cm_mosaic_grid}. In adopting this approach, a balance between the sample size and its ‘purity’ is paramount. We discern prominent clusters through a adjusted median split for each metric. We define Upper and Lower thresholds by adding and subtracting a 15\% factor to the median to establish the bounds for cluster selection per metric. Our algorithm picks the most prominent clusters for each metric, crossmatches these selections, and collates them. From this, we discern the prominent clusters representing Spiral and Elliptical galaxies. Figure \ref{fig:selection_metrics} illustrates this process, displaying the \texttt{EGG} system distributions with the applied cuts. It visually demarcates the preserved data segments (outside the black lines) and the \texttt{T-Type} range for the galaxies.

\subsubsection{Grid size and distance}

\begin{figure*}
	\includegraphics[width=505px]{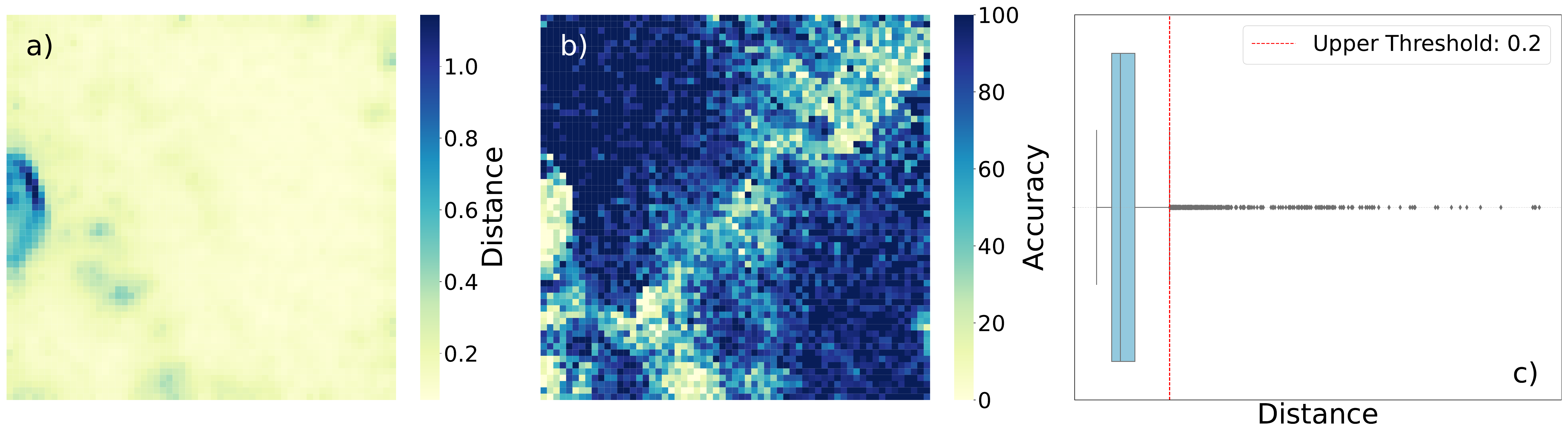}
    \caption{Panels a and b showing the cluster grid (60$\times$60). Both of these panels show the interplay between distance and accuracy. An additional selection criterion is the inter-cluster distance. Panel a visualizes the distance between clusters and their neighbors, while Panel b illustrates accuracy per cluster (similarly to Figure \ref{fig:cm_mosaic_grid}). A noteworthy observation is the direct correlation between greater distances and decreased accuracy. Panel c identifies outliers in distances, which are subsequently excluded from prominent clusters.}
    \label{fig:selection_distance}
\end{figure*}

Although we previously utilized a 4x4 grid for illustrative clarity, it is essential to adjust to an optimal grid size before deploying the selection algorithm. Per SOMbrero documentation \citep{vialaneix2022package}, the ideal grid size aligns with the dataset's volume and can be formulated as:
\[ \text{grid\_size} = \sqrt{n\_subjects \times 0.1}. \]
For our dataset comprising 36,000 galaxies, the optimal grid dimensions gravitate towards 60$\times$60, yielding 3,600 clusters. Henceforth, our analyses 
adhere to an optimal grid configuration of 60$\times$60.

Despite being identified as prominent through metric-based criteria, not all clusters guarantee high classification accuracy. Some might display a mix of galaxy types due to factors like segmentation anomalies or intrinsic galaxy irregularities. To circumvent such complexities, we further refine our cluster selection by evaluating the average inter-cluster distance. Generally, clusters with more substantial average distances indicate irregularities, often correlating with diminished accuracy.

\begin{figure}
	\includegraphics[width=\columnwidth]{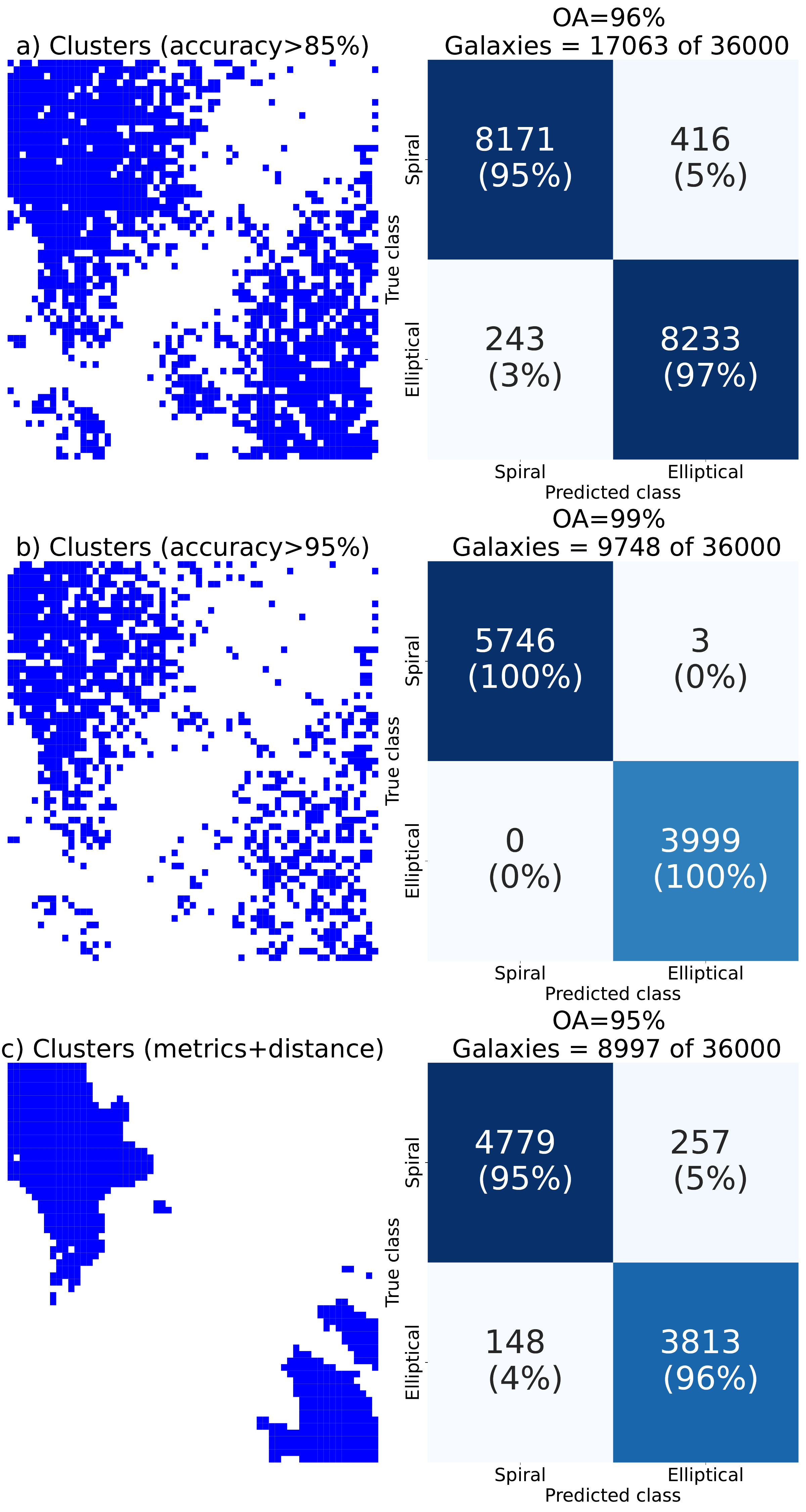}
    \caption{A comparative snapshot of cluster selection outcomes. Rows 1 and 2, serving as an example to SOMbrero's prowess in class differentiation based on morphological metrics, utilize accuracy thresholds (85\% and 95\%). These rows provide insights into the highest accuracy clusters, pinpointing their spatial distribution and the resultant confusion matrix, had we exclusively chosen them. Row 3 delves into our method's outcomes, integrating both prominent morphological criteria and distance considerations.}
    \label{fig:selection_results}
\end{figure}

In the context of Self-Organizing Maps (SOMs), each grid node is intrinsically tied to a weight vector mirroring the input data's dimensionality. 
The fundamental objective of SOM training is to iteratively adjust these weight vectors such that they approximate the input data vectors. This allows the SOM to create a spatial organization of the input data in a lower-dimensional space. A critical aspect of SOMs is the calculation of distances between the weight vectors of different nodes (clusters) in the SOM grid. For this study, we compute these distances using the Letremy distance \citep{fort2002advantages}, quantifying the dissimilarity between nodes. Following the training process, nodes with similar weight vectors—and thus closely located in the SOM grid—are considered to form groups. Therefore, the distance between clusters and their neighbors refers to the dissimilarity between these groups of nodes. This represents how distinct one cluster is from its neighboring clusters based on their representative features.

The inter-cluster distances and their proximal neighbors find visualization in Figure \ref{fig:selection_distance}. 
Notable
correlation is shown in this figure between the neighboring distances (shown in panel (a)) and the accuracy of individual clusters (depicted in panel (b)). The metric of distance is acutely sensitive, and a mere elimination of outliers efficiently sieves out low-accuracy clusters, evidenced in panel (c).

In order to increase the number of selected galaxies and make the selection process more robust, we employ an ensemble selection methodology. The procedure involves constructing the grid and selecting galaxies from prominent clusters three times. Each time, the grid has a unique size. As previously mentioned, the ideal grid dimensions are computed as \( \text{grid\_size} = \sqrt{n\_subjects \times 0.1} \pm 1 \). Running the procedure just once would involve using a \(60\times60\) grid and selecting galaxies from prominent clusters. Instead, we generate the grid with dimensions \( \text{ideal\_grid\_size} \pm 1 \), i.e., with sizes \(59\times59\) and \(61\times61\). This ensemble approach produces predictions of a galaxy's label across three different grids. We select all galaxies located within the prominent clusters across these grids, ensuring the elimination of duplicates. This technique increases the galaxy count by approximately 15\% without impacting the overall accuracy (OA) compared to a singular grid selection.
To surmise, our prominent cluster selection rides on the back of the morphological metrics, and the average distance from neighboring clusters. The described method is intuitive, with results demonstrated in Figure \ref{fig:selection_results}, culminating in clusters possessing a commendable OA. Still, we recognize the potential horizon for refinement in the selection dynamics.

In Figure \ref{fig:selection_results}, a tri-row visualization provides a deep dive into the nuances of prominent cluster selection. Rows 1 and 2 portray cluster distribution for two scenarios: one filtering clusters exceeding 85\% accuracy and the other breaching the 95\% threshold. The confusion matrices offers a snapshot of the resulting OA if these specific clusters were selected from the grid. It is worth noting that SOMbrero exhibits a faculty for grouping galaxies in clusters with 85\% or superior accuracy. For instance, a focus on these galaxies alone returns an OA of 96\%. Venturing into clusters with an accuracy beyond 95\% nudges us closer to an impeccable classification. The results showcased in rows 1 and 2 are primarily for demonstrative purposes. In an unsupervised learning paradigm, access to accuracy metrics beforehand is not available. Nevertheless, these illustrations are illuminating, reinforcing the prowess of morphological metrics in delineating galaxy classes proficiently.

When it comes to our method of prominent clusters selection, as depicted in row 3, the OA for the selected galaxies registers at 95\%. This infers that the predominant fraction of selected clusters aligns with an accuracy threshold akin to 85\%. Intriguingly, we pinpoint a more pronounced misclassification rate for Spiral galaxies in contrast to their Elliptical counterparts. This trend is congruent with the inherent complexities of Spiral galaxies, which sport varied and potentially intricate configurations that might pose difficulties when distilled through morphological metrics alone. A salient observation emerges when considering the dataset's size. The foundational dataset channeled into the SOMbrero comprised 36,000 galaxies. However, in the selection scenarios articulated, a tangible decrement in these numbers becomes evident, given our focus on selectively curating only the 'prominent' clusters. To quantify, the proportions of the original galaxy count retained across the scenarios are 47\%, 27\%, and 25\%, respectively. On a broader scale, we observe a contraction by approximately three times during the selection phase. This underscores the indispensability of ample dataset volumes to harness the full potential of this method. Concretely, as the dataset's magnitude swells, the efficacy and utility of our approach proportionally intensify.

Figure \ref{fig:selection_results} demonstrates the juxtaposition between the performance metrics of the entire grid versus those of the prominent clusters. The fruits of our algorithmic endeavors yield predicted labels for prominent selection. Specifically, the matrix illuminates the discrepancies between the predicted and true classifications, offering a quantitative measure of the algorithm's accuracy.

Several noteworthy observations can be distilled from our analysis:

1. \textbf{Morphological Metrics as Effective Descriptors:} The results accentuate the potency of morphological metrics as robust numerical descriptors. They manifest an impressive capability to discern between diverse morphological galaxy classes.

2. \textbf{Utility of Unsupervised Machine Learning:} The employment of unsupervised machine learning, particularly the SOMbrero package in our research, emerges as a dependable and effective approach for curating ground-truth galaxy catalogs. Such catalogs bear immense potential, standing as viable alternatives or enriching supplements to citizen science endeavors like Galaxy Zoo.

With this foundational understanding secured, we are poised to explore practical applications of this catalog. Specifically, our subsequent steps
involve employing this catalog as a ground-truth basis for the training of a supervised model. Once trained, this model is tasked with classifying a larger and more comprehensive galaxy catalog. Its performance is rigorously evaluated through comparisons with the true labels. For the remainder of this study, we refer to this sample as the 'training catalog'.

This concludes the first component of our proposed method. If our work were to halt at this juncture, we would be left with a classified catalog of 36,000 galaxies, each assigned with varying degrees of accuracy, as well as a SOMbrero model proficient in classifying input based on morphological metrics. Our next phase, however, is to leverage the galaxies from the prominent clusters identified in this study as input data for a supervised deep learning paradigm.

\subsection{Supervised Machine Learning}

Our resulting catalog, derived from the SOMbrero unsupervised classification, is notably constrained in size, limiting its standalone applications. Empirical tests with our method affirm that producing a sizable unsupervised classification would incur a prohibitively extensive computational time cost. Yet, the need remains for an effective method to extend our results to larger datasets, thereby enhancing the utility and impact of our work. To achieve this scale, we propose the application of Supervised Machine Learning (SML) techniques.

\begin{figure}
	\includegraphics[width=\columnwidth]{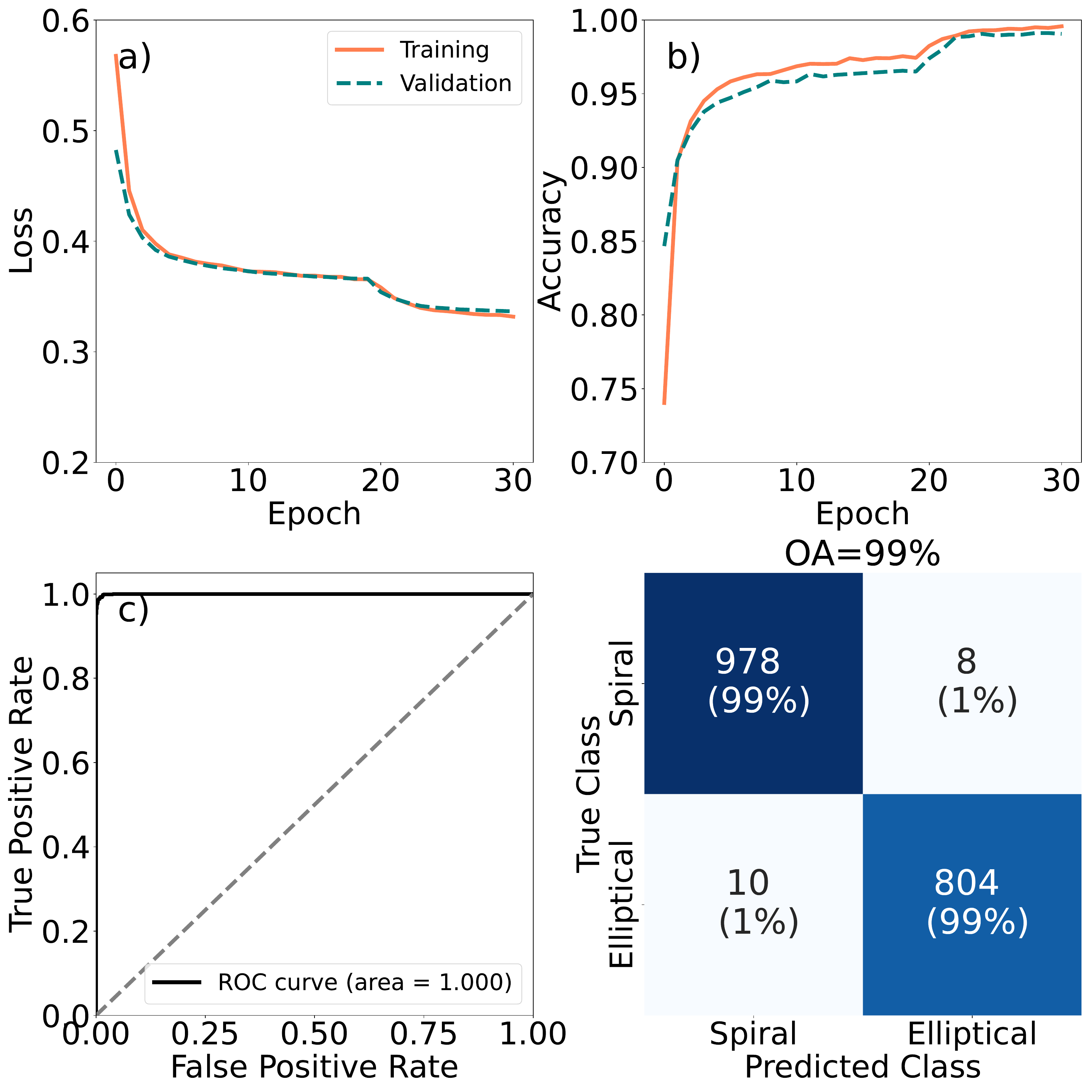}
    \caption{The supervised model training report.  Panels (a) and (b) display the training curves of Loss and Accuracy, respectively. The curves in both panels reflect a healthy model behavior without overfitting or underfitting. Panel (c) illustrates the Receiver Operating Characteristic (ROC) curve, which demonstrates an almost perfect ability of the model to discern between the two classes. Panel (d) shows the CM with OA. This panel not only presents the accuracy on the test dataset but also provides the number and fractions of galaxies that were correctly and incorrectly classified.}
    \label{fig:supervised_analisys}
\end{figure}

In its current state, the training catalog has two issues. Notably, it is relatively small (compared to the test data) and is not entirely free from noise. Specifically, our dataset includes instances where Spiral galaxies have been mislabeled as Elliptical galaxies, and vice versa. Through our prominent cluster selection process, we have constrained this noise to approximately 2\% for Elliptical galaxies and 6\% for Spiral galaxies. Nonetheless, this noise represents a critical concern: it must be addressed to ensure our model can both learn effectively and generalize accurately to unseen data. 
The efficacy of deep CNNs has been extensively validated in various studies \citep{ dominguez2018improving, barchi2020machine, khalifa2017deep, primack2018deep, khan2019deep, tohill2021quantifying, walmsley2022galaxy, cheng2023lessons}; these networks are capable of training on labeled data and subsequently applying this learning to classify, predict, or estimate values in much larger, analogous datasets. Our study adopts a similar approach, with a distinguishing feature: our employment of a training catalog constructed via an unsupervised machine learning (UML) algorithm that is grounded in morphological metrics. This strategy integrates the best of both UML and SML paradigms. It capitalizes on the UML’s capacity to generate a high-quality, albeit smaller, labeled dataset and marries this capability with the SML’s proficiency in scaling this learning to significantly larger datasets. Through this synergy, we aim to construct a robust, scalable, and highly accurate galaxy classification model while concurrently minimizing the time and computational resources required. Furthermore, we leverage the benefits of Transfer Learning (TL) to enhance the performance of our model and decrease the training time. Transfer Learning is an approach that focuses on utilizing knowledge acquired from one problem (source domain) to address a related, yet distinct problem (target domain). This strategy effectively circumvents the 'cold-start' issue prevalent in machine learning—the requirement to learn each new task from scratch, which often necessitates extensive computational resources and large volumes of labeled data \citep{pan2009survey}. 

In this study, we employ a popular form of Transfer Learning that involves using pre-trained deep learning models. These models, trained on large-scale datasets, are repurposed for a specific task by replacing and fine-tuning the last few layers of the network. The underlying assumption of this approach is that earlier layers of deep networks learn general, reusable features, while later layers become increasingly task-specific. Our methodology incorporates a transfer learning approach, serving two main objectives: reducing the necessary data volume for effective training and expediting the training process by limiting the required epochs. We utilize pre-trained weights from the ImageNet dataset in this study \citep{deng2009imagenet}. These weights, optimized to discern a plethora of features in the source domain, are aptly modified in our model to classify galaxies based on their morphological distinctions. With these pre-trained weights as a foundation, our model undergoes fine-tuning on our specific dataset during training. This retains the high-level features learned from ImageNet while adapting to our galaxy classification task's specific intricacies. As a result, our model achieves commendable accuracy using a significantly smaller training dataset in a condensed timeframe, underscoring the efficacy of the transfer learning paradigm.

For our pipeline, we have chosen TensorFlow as our framework. This open-source machine learning framework, developed by \citet{abadi2016tensorflow}'s Brain Team, offers a comprehensive ecosystem that facilitates the development and implementation of a wide range of machine learning models and complex numerical computations. TensorFlow operates through data flow graphs where nodes represent mathematical operations and edges denote multidimensional data arrays—known as tensors—that flow between these nodes. For our primary model architecture, we employ Xception, a deep learning model 
"Xception," standing for "Extreme Inception," is a modification of the Inception architecture. Its primary innovation lies in the use of depthwise separable convolutions, instead of the standard convolutions found in Inception. Comprising 36 convolutional layers, Xception forms the feature extraction base of our model \citep{chollet2017xception}.

\subsubsection{Training Supervised model}\label{training_super_model}
The supervised learning pipeline in our study adheres to a conventional approach, complemented by techniques specifically designed to mitigate the impact of noise in the data. The labeled catalog, derived from our unsupervised clustering model (SOMbrero), serves as the basis for our supervised learning stage. We partition this labeled dataset into distinct subsets, allocating 80\% for training, 20\% for validation. The input data comprises RGB images, sourced directly from the Sloan Digital Sky Survey (SDSS) website. Aside from the standard preprocessing steps requisite for the input to our chosen model, we apply no additional preprocessing to these images.

To address the issue of noise in our dataset, we employ two strategic techniques: (1) label smoothing, and (2) ensemble training.

1. \textbf{Label Smoothing}: In the realm of deep learning, label smoothing is a regularization technique aimed at reducing the model's confidence in the assigned labels of the training dataset. This is implemented to enhance the model's generalization performance. Traditionally, when training a deep learning model with a Binary Cross-Entropy loss function, the target labels for a binary classification task are designated as "1" for the positive class and "0" for the negative class. Label smoothing seeks to moderate this clear-cut, binary assignment. Instead of using a "1" for the positive class, it employs a slightly reduced value (in our case, empirically deducted, is 0.8), and conversely, replaces the "0" for the negative class with a slightly increased value (in our case, empirically deducted, is 0.2). This nuanced approach is designed to prevent the model from developing excessive confidence in the training data, which can be vital for minimizing overfitting and noise.

2. \textbf{Ensemble Training}: Another effective technique to counteract the effects of noise is ensemble training. This method involves training multiple models on identical data but with different random selections for the training, validation, and test datasets. By using diverse subsets of the data for each individual model in the ensemble, we introduce a level of variation that can decrease the chance of overfitting and improve the robustness of the overall predictive performance. This ensemble strategy aims to leverage the strengths of multiple models, producing more stable and reliable predictions less influenced by the noise in the training data.

In Figure \ref{fig:supervised_analisys}, we present a comprehensive analysis of the model's training and testing performance, highlighting the effectiveness of these techniques in practice. Figure \ref{fig:supervised_analisys} showcases the consistent, stable behavior demonstrated by our model throughout the training process, as well as the high accuracy achieved in the testing phase. The impressive performance of the model suggests that the architecture of our neural network aligns well with the unsupervised labels produced by the SOMbrero algorithm. Specifically, this outcome may indicate that our neural network architecture excels at identifying discriminative features that effectively differentiate between the two galaxy classes — Elliptical and Spiral. 
Owing to the variable noise levels in each data slice, individual models manifest differing mislabeling degrees. However, this is effectively counterbalanced by our ensemble strategy. In our ensemble methodology, each model classifies the final dataset independently. Their resultant classification probabilities are then amalgamated using a simple mean. This averaging process neutralizes errors across models, resulting in a more stable and accurate final classification.

\begin{figure}
	\includegraphics[width=\columnwidth]{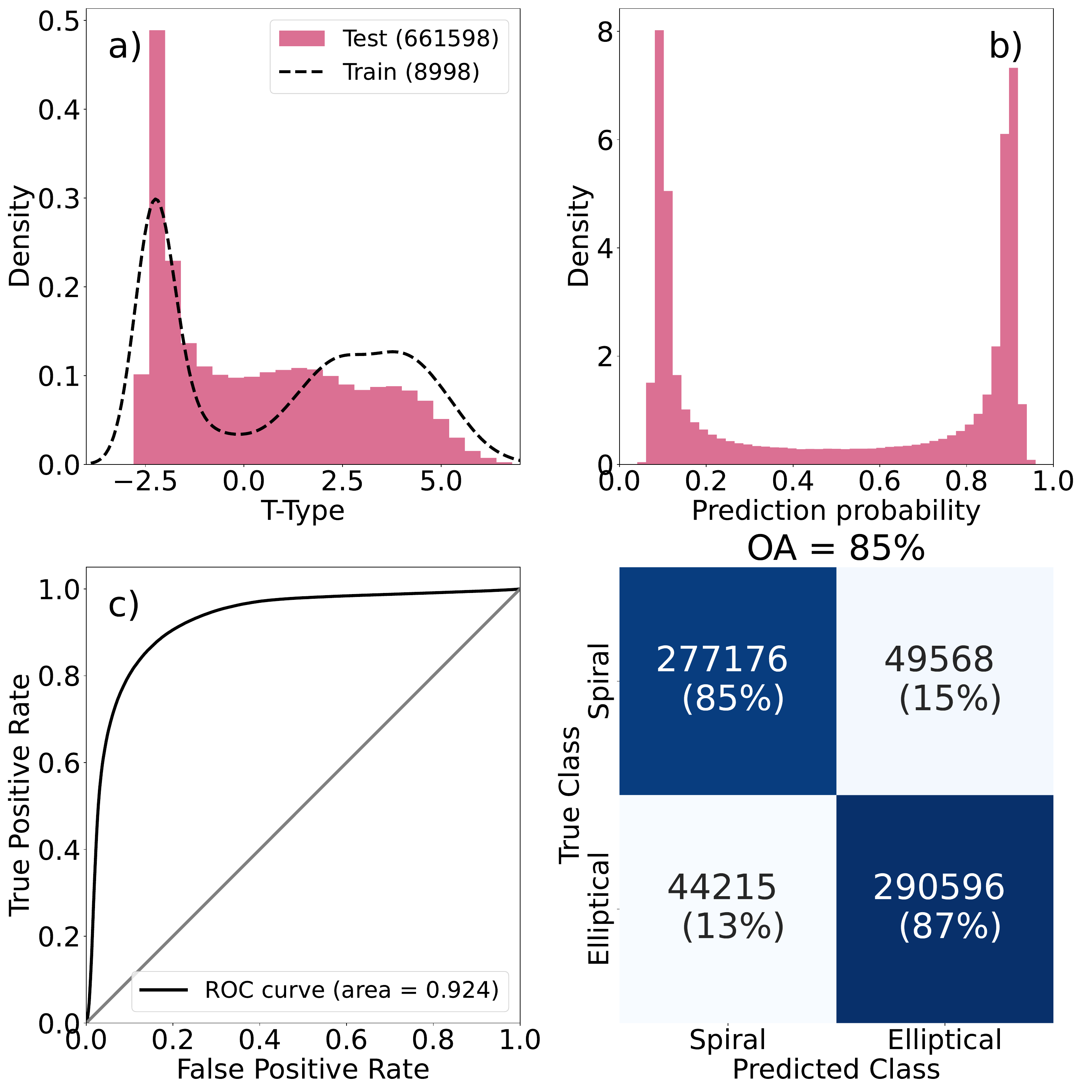}
    \caption{Data selection (a), prediction probabilities (b), ROC curve (c), and confusion matrix result (d) when applying the trained model (without retraining) to the remaining data from the \citet{dominguez2018improving} catalog, as delineated in Figure \ref{fig:data_combined}.}
    \label{fig:test_dominguez_ensemble}
\end{figure}

Our training process commences by loading the pre-trained model from the Keras database, excluding the final output layer. In its place, we introduce our custom binary classification layer, designed to discern between two primary classes of galaxies — Spiral and Elliptical. Our approach to training the model is bifurcated into two distinct phases: (1) Feature Extraction (FE) and (2) Fine Tuning (FT).

During the initial \textbf{Feature Extraction (FE)} phase, we intentionally "freeze" the convolutional base that was created from the pre-trained model. Specifically, we set \textit{layer.trainable = False} for all layers in the convolutional base, effectively preventing the weights in these layers from being updated during this training phase. The primary objective of this pivotal step is to acclimatize our model to the new data without perturbing the weights that have already been learned from the pre-trained model. To achieve this, we exclusively train the few top layers that we appended to the pre-trained model.

\begin{figure}
	\includegraphics[width=\columnwidth]{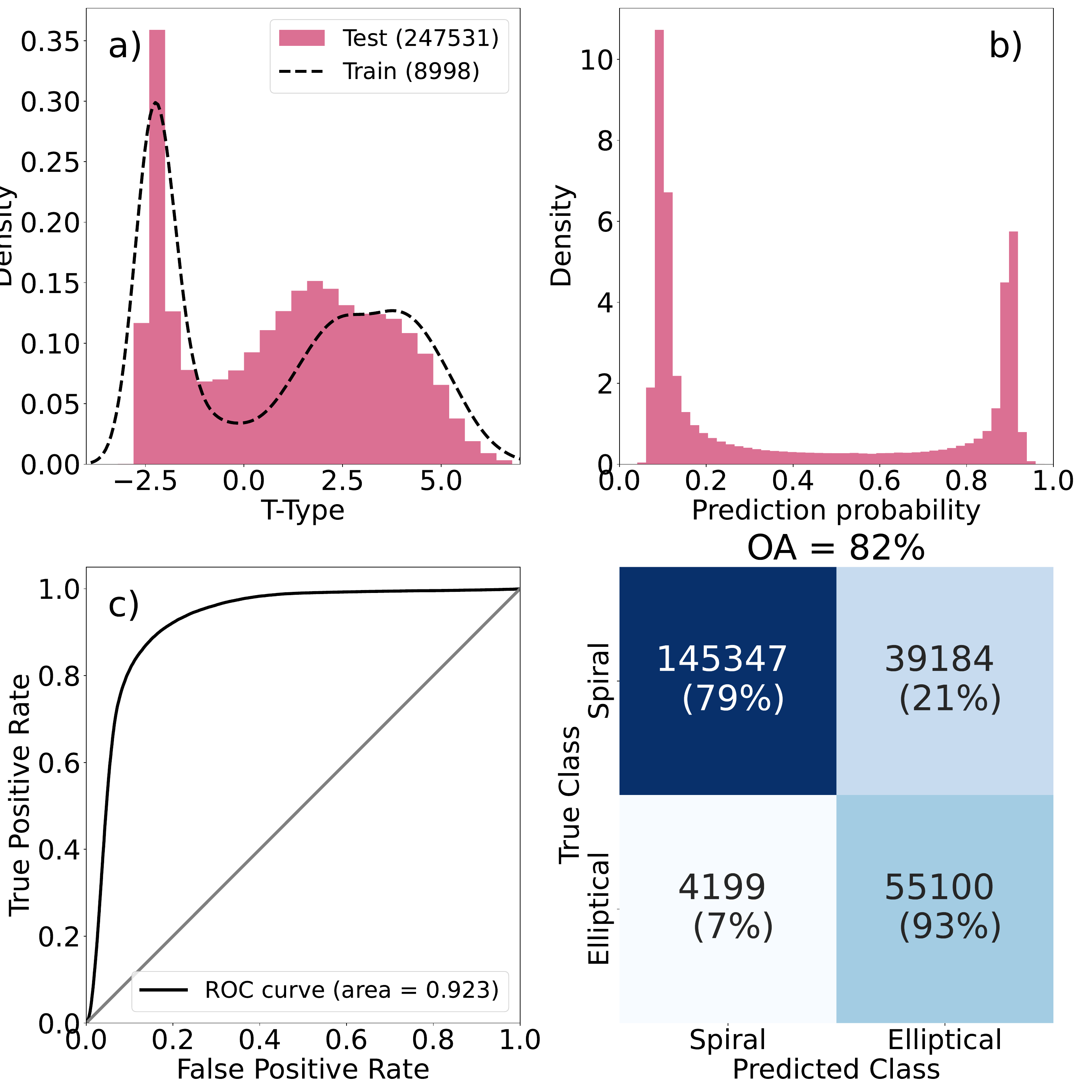}
    \caption{Data selection (a), prediction probabilities (b), ROC curve (c), and confusion matrix result (d) of applying the trained model (without retraining) on the Galaxy Zoo 1 (GZ1) data.}
    \label{fig:test_gz_1}
\end{figure}

\begin{figure*}
	\includegraphics[width=505px]{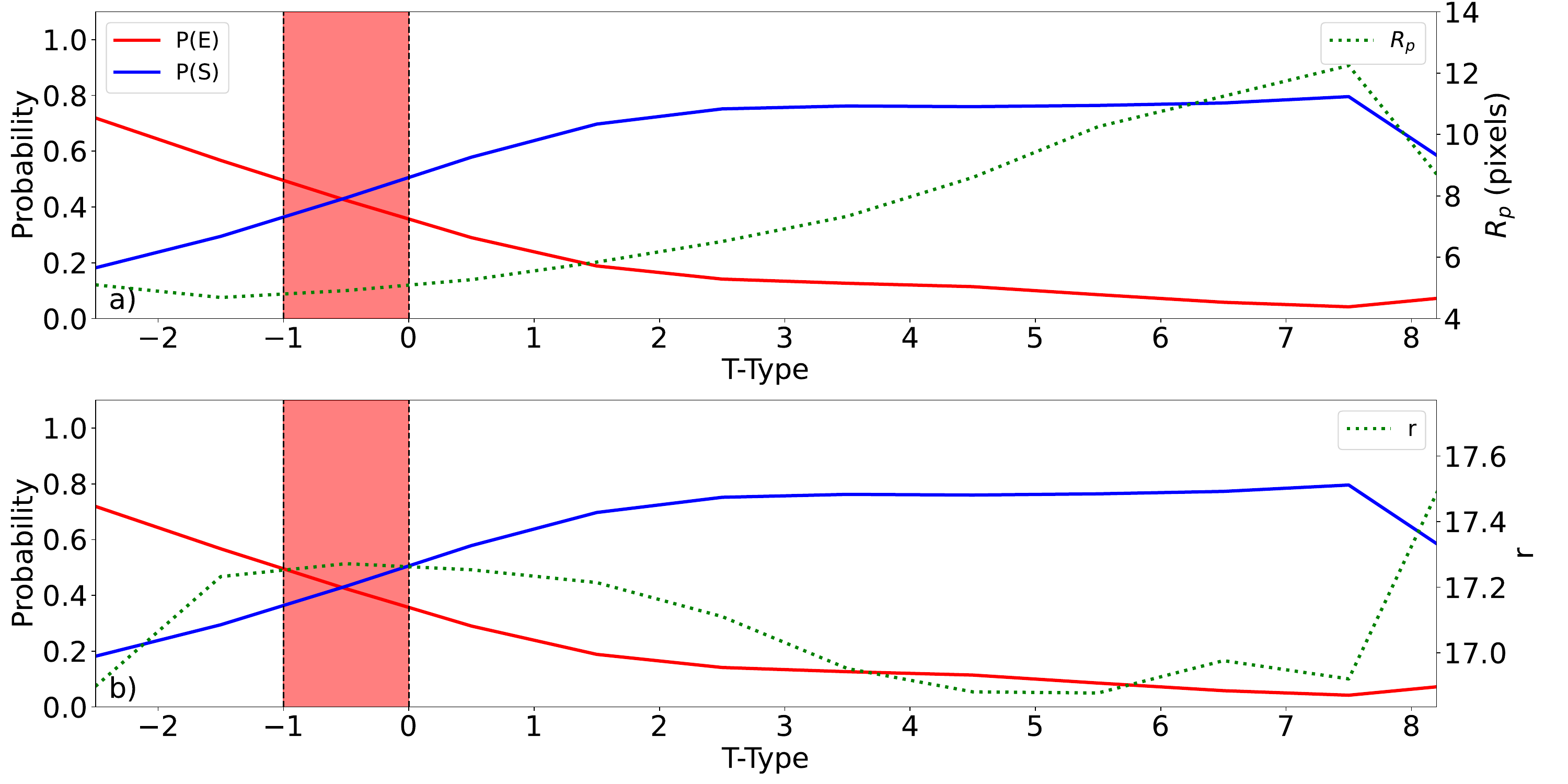}
    \caption{Debiased probability of galaxy being Elliptical (P(E)) and Spiral (P(S)) from GZ1 \citet{lintott2011galaxy} plotted against \texttt{T-Type} ranges from \citet{dominguez2018improving}. Panel (a) plots petrosian radius ($R_p$), while panel (b) plots magnitude (r). Red area between  $-1 < \texttt{T-Type} < 0$ represents galaxies that to be removed from the testing catalog due to low confiability of truth class.}
    \label{fig:critical_ttype}
\end{figure*}

Figure \ref{fig:supervised_analisys}, panels a and b, demonstrate this process graphically. Notably, a steep slope is observable on both the training and validation curves. This juncture demarcates the transition from the FE phase to the FT phase. Epochs one through twenty are designated to the FE stage, during which we observe a slow, methodical learning process. The primary aim at this stage is to establish an initial association between the weights in the newly added layers and the galaxy data. Subsequently, during the \textbf{Fine Tuning (FT)} phase, epochs 20 through 30, we decrease the learning rate and selectively "unfreeze" additional layers within the model, thereby allowing these layers to be trainable. This strategic alteration enables the model to refine and adjust the weights based on the specificities of our galaxy dataset, thereby enhancing its predictive capabilities. This critical transition is illustrated in Figure \ref{fig:supervised_analisys}, at the point where the steep slope is evident. At this juncture, we witness a marked reduction in loss and a concomitant increase in accuracy, signalling the effective fine-tuning of our model. The Fine-Tuning (FT) phase is initiated only subsequent to the completion of the training of the uppermost classifier layers, while the pre-trained base model remains in a non-trainable state. This sequence is of paramount importance. If one were to append a randomly initialized classifier to a pre-trained base model and attempt to train all layers in unison, the gradients during the backpropagation could potentially be of substantial magnitude. This scenario arises due to the randomly initialized weights of the appended classifier, which are initially uncorrelated with the learned features of the pre-trained base. Such large gradients, propagating through the network during the training phase, can inadvertently cause a significant alteration of the more generalized and useful features learned by the base model during its training on the initial dataset. This phenomenon is colloquially termed as the pre-trained model 'unlearning' its valuable, previously acquired knowledge. Thus, the FT phase is strategically designed to circumvent this issue. Initially, the pre-trained base model is frozen (non-trainable), allowing the appended classifier to become adequately trained with a stable set of weights. Only subsequent to this, the fine-tuning phase is embarked upon. During this phase, selected layers of the pre-trained model are unfrozen and are gradually adapted to the new data, under much smaller learning rates. This ensures that the updates to the weights are incremental and controlled, preserving the integrity of the learned features in the pre-trained base model.

This delineates the culmination of the model's training phase. The training uses labels that were generated during the unsupervised learning phase. The ensuing step in our methodology involves deploying this trained model to classify an expansive, remaining dataset, which encompasses over half a million galaxies. This is succeeded by a thorough analysis of the resultant classifications, thereby assessing the efficacy and generalization capability of our model on a substantially larger and potentially more diverse dataset.

\subsubsection{Testing model - \citet{dominguez2018improving} catalog}

After the completion of the model’s training phase, we proceed to the large-scale application of this trained model, exploiting its ability to classify a significantly bigger dataset. This stage is dedicated to the classification of the remaining 661,598 galaxies, leveraging the model trained on the dataset labeled through the unsupervised learning algorithm. To ensure the integrity and independence of this phase, meticulous care has been taken to ensure that the images in this final test dataset are entirely novel and have never been exposed to the model during its training or validation phases. This is achieved by removing all the galaxies that were part of the dataset used for the supervised training step. This ensures a fair assessment of the model's generalization capability on unseen data, avoiding any data that could artificially inflate the model’s performance metrics.

Figure \ref{fig:test_dominguez_ensemble} illustrates the results of applying the trained ensemble model on the final dataset comprising 661,598 galaxies. It begins with data distribution which shows the \texttt{T-Type} range, comparing the data used for training with the testing dataset. Following that, the ROC curve is a graphical plot that illustrates the diagnostic ability of the model as its discrimination threshold is varied. 
The area under the ROC curve (AUC) can be used as a single scalar metric to summarize the model's ability to discriminate between the two classes (Spiral and Elliptical) effectively. A higher AUC suggests that the model has a high capacity to distinguish between Elliptical and Spiral galaxies, affirming the model's strong performance on the unseen dataset. As for Panel c, the majority of galaxies in both classes appear to be correctly classified, as indicated by the significant values along the diagonal of the matrix. The OA, which represents the proportion of total instances that were classified correctly, indicates that the model is effective in classifying galaxies into their respective morphological classes based on the unseen data. The Confusion Matrix further allows for the inspection of class-specific performance, shedding light on whether the model has a bias toward a particular class.

From this analysis, we can conclude that the model was able to learn and generalize in order to be successfully applied to the much larger dataset (almost 30 times bigger than the training sample). Moreover, the model is capable of replicating galaxy morphology derived via \texttt{T-Type}. 
The mislabeling of both classes is similar, which indicates that the model is equally competent in discerning galaxy morphology.

\subsubsection{Testing model - GZ1 catalog}

Following the primary testing phase, the trained model was further assessed using data from the GZ1 project. As detailed in Section \ref{data_processing}, from the comprehensive GZ1 catalog, we selected galaxies that exhibit a classification certainty exceeding 80\% (excluding those flagged as \textit{UNCERTAIN}). The results are depicted in Figure \ref{fig:test_gz_1}.

When juxtaposed with outcomes derived from the residual catalog of \citet{dominguez2018improving}, our model's accuracy was similar in case of OA and class specific accuracy. This indicates a correlation between human-based assessments and morphological metric classifications, as well as \texttt{T-Type} value. Furthermore, the concordance between human-annotated labels from Galaxy Zoo and morphological metrics endorses the efficacy of our approach. It suggests that well-trained automated techniques can closely mirror human perceptions in identifying intricate patterns in galactic morphologies.

\begin{figure}
    \includegraphics[width=\columnwidth]{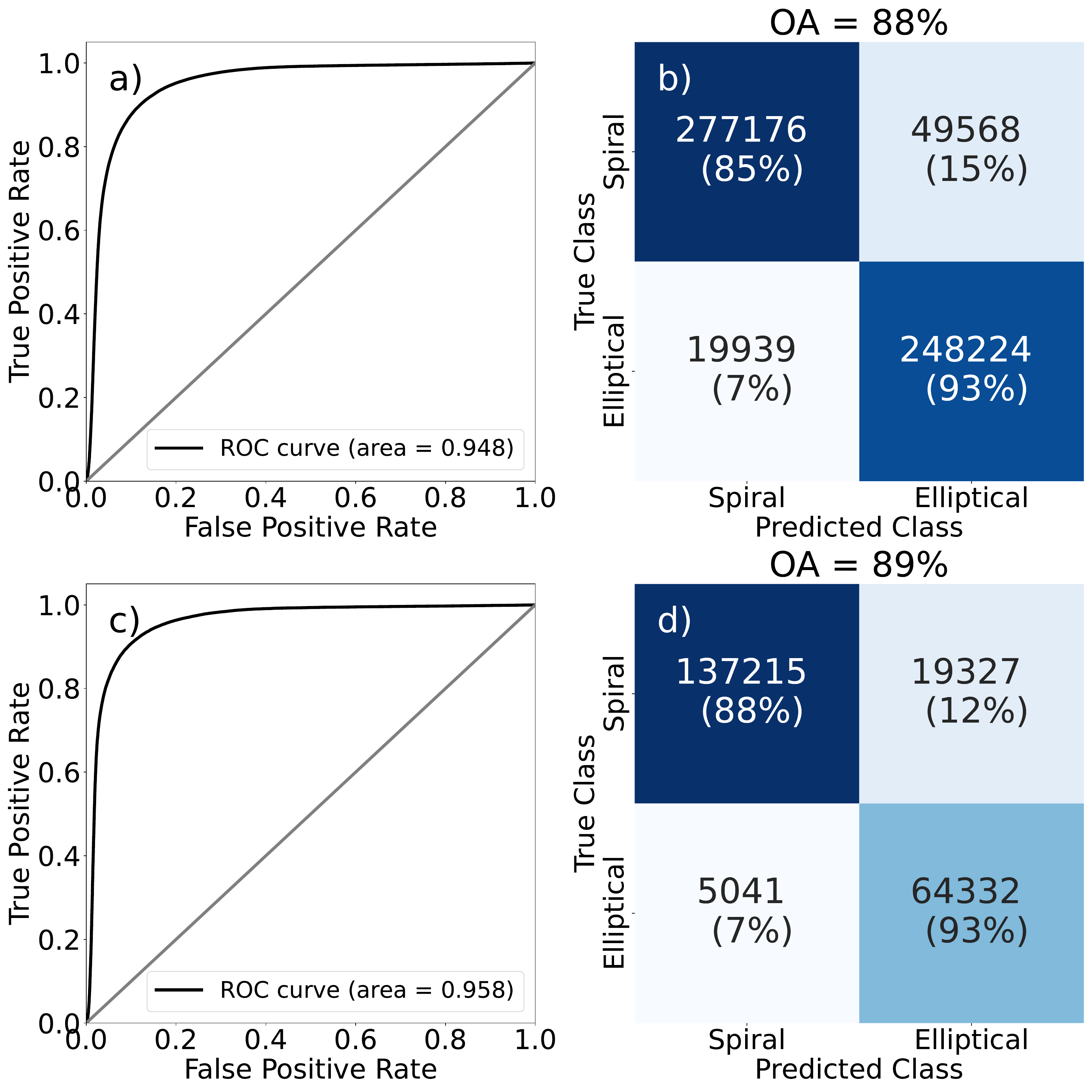}
    \caption{Performance assessment of the previously trained model on dataset cleared from ambiguous galaxies reported in Figure \ref{fig:critical_ttype}. Top row shows results for \citet{dominguez2018improving} and bottom row shows results for \citet{lintott2011galaxy} catalog.}
    \label{fig:test_10_ensemble}
\end{figure}

\subsubsection{\texttt{T-Type} and bias}\label{ttype_bias}

When analyzing data, it is imperative to acknowledge its limitations and devise strategies to address or circumvent them. Notably, no catalog of significant magnitude can claim to be devoid of biases or guarantee a 100\% accuracy rate in classifications. This observation is particularly pertinent to expansive catalogs such as that by \citet{dominguez2018improving} or \citet{lintott2011galaxy}. A substantial portion of the galaxies in this catalog has not undergone visual verification. Furthermore, given that \texttt{T-Type} represents continuous values, there exists a juncture where classes overlap, leading to inherent ambiguities and an elevated likelihood of misclassification. As a result, it is not feasible to evaluate a new methodology on the entire dataset without excluding ambiguous or questionable entries. This brings forth the essential discussion of identifying galaxies that may be omitted from the testing catalog to ensure a sample that provides a robust and trustworthy foundation for testing our proposed method.

In Figure \ref{fig:critical_ttype}, we depict the debiased probability of each class from \citet{lintott2011galaxy} plotted against \texttt{T-Type} ranges as documented by \citet{dominguez2018improving}. Additionally, panel (a) illustrates $R_p$, and panel (b) displays Magnitude. Several crucial insights emerge from this representation. A primary observation is that galaxies confined within the range $-1 < \texttt{T-Type} < 0$ confront the most significant classification ambiguity. They lie at the juncture of overlapping classes, making their definitive classification challenging. Adding to the complexity, these galaxies are both the smallest in size and faintest in magnitude, reinforcing the difficulty in their classification. Consequently, it becomes imprudent to include them in the ground truth catalog. Therefore, we evaluate our previously trained model (presented in subsection \ref{training_super_model}) using subsamples from which galaxies in the interval $-1 < \texttt{T-Type} < 0$ have been removed. This approach aims to eliminate potential 
extra variance
in the ground-truth labels, which could otherwise negatively impact the model's performance. The model was not retrained or modified; the only difference lies in the testing catalogs used, specifically those from \citet{dominguez2018improving} and \citet{lintott2011galaxy}.

In Figure \ref{fig:test_10_ensemble}, we present the outcomes of our model trained on datasets from which galaxies exhibiting ambiguity, as highlighted in Figure \ref{fig:critical_ttype}, have been excluded. A significant observation is the marked improvement in the OA for the GZ1 catalog. This enhancement is even more pronounced for the Spiral class. Notably, the model, when trained on labels derived from unsupervised clustering, predicted galaxies within the $-1 < \texttt{T-Type} < 0$ range as Elliptical, aligning with intuitive expectations. In contrast, labels from GZ1 suggested otherwise.

\begin{figure}
	\includegraphics[width=\columnwidth]{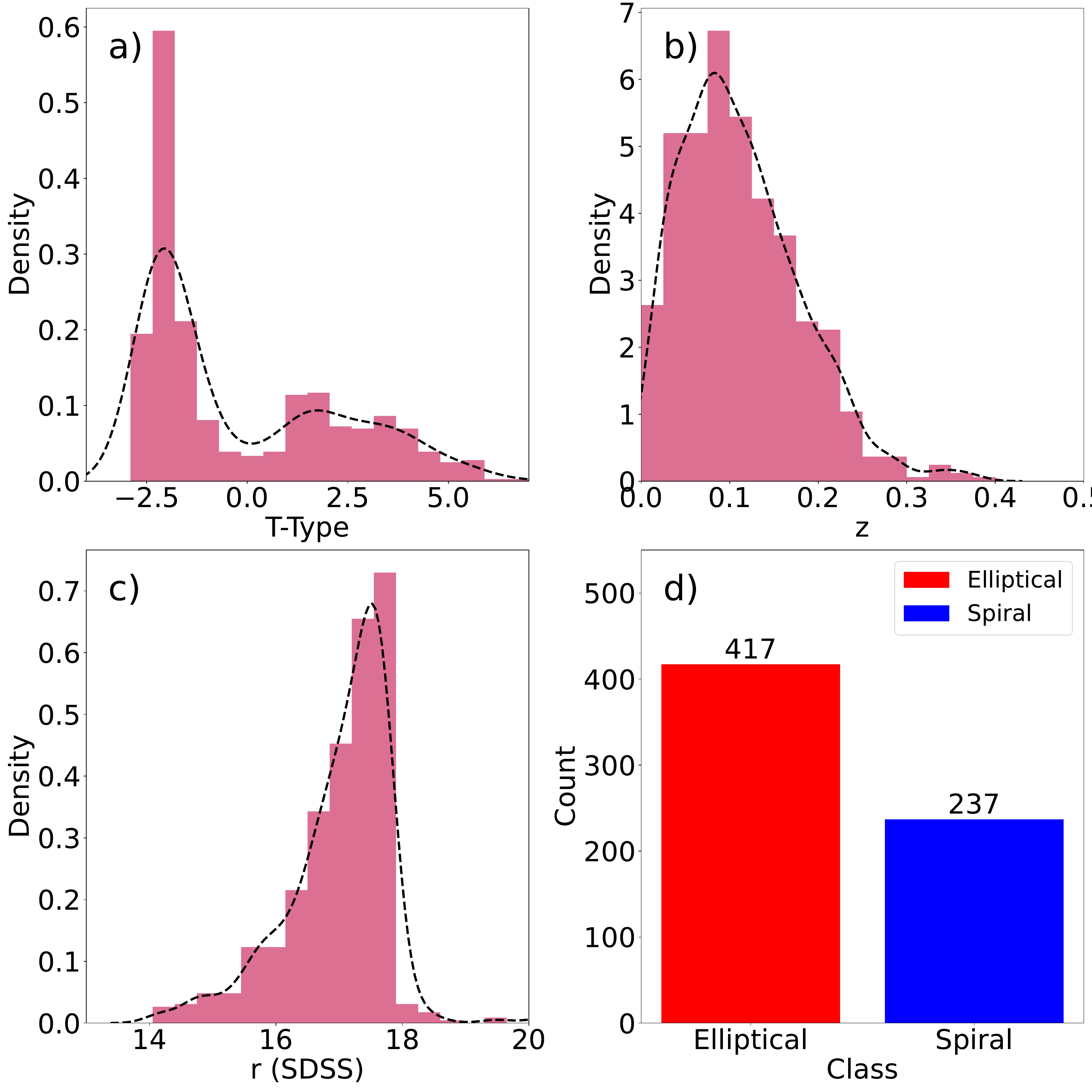}
    \caption{Illustrative representation of the assembled HST catalog. Panel (a), Panel (b), and Panel (c) depict the distributions of morphological \texttt{T-Type}, redshift, and magnitude, respectively. Panel (d) quantifies the relative abundances of Elliptical and Spiral galaxy classes within the catalog.}
    \label{fig:sdss_vs_hst_fits}
\end{figure}

\begin{figure*}
	\includegraphics[width=505px]{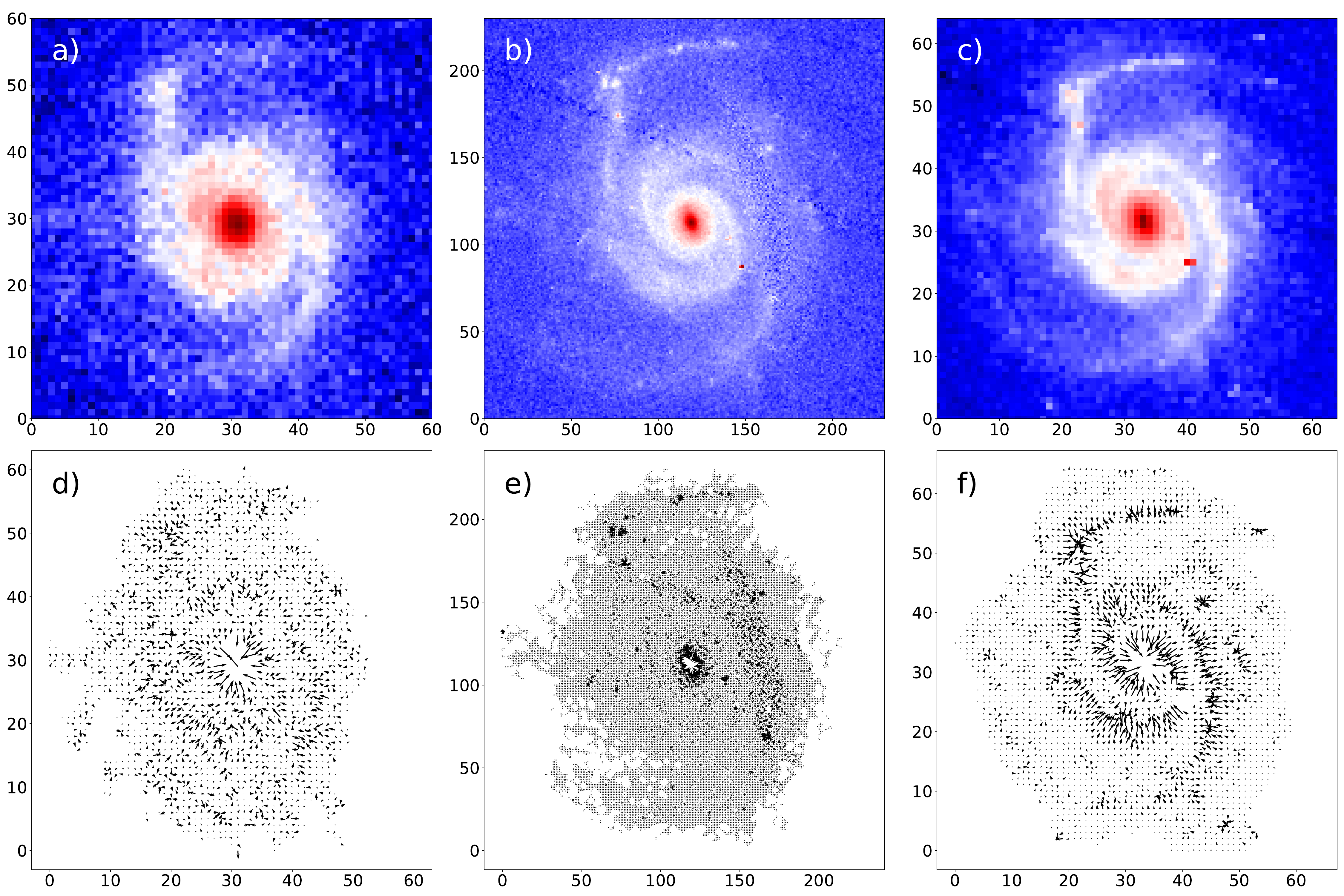}
    \caption{Comparative visualization of a galaxy image (top row) and galaxy gradient (bottom row) as observed by SDSS (Panels a, d) and HST (Panels b, e). Panels (c) and (f) depicts a downsampled version of the HST image to the same resolution of SDSS for reference.}
    \label{fig:sdss_vs_hst_ds_fits}
\end{figure*}

Examining the results on the dataset from \citet{dominguez2018improving}, we observe a contrasting situation, consistent enhancement in classification performance in Elliptical galaxies, complemented by a moderate increment in OA. This observation mostly aligns with the anticipated outcome that eliminating ambiguous galaxies would bolster the model's overall performance.

Drawing upon these findings, it is plausible to infer that our model exhibits the capacity to navigate and interpret intricate datasets, aligning its classifications closely with \texttt{T-Type}. We argue that the metrics are inclining towards a concurrence with \texttt{T-Type} classifications, suggesting that they offer a more robust and consistent framework, primarily since they are grounded on composite base on human-driven visual classifications along with highly curated \citet{nair2010fraction} catalog.
\section{Extension of our Methodology to HST Data}\label{section_sdss_hst}
In this section, we introduce an extension of the method to Hubble Space Telescope (HST) data. Although the SDSS provides coverage over vast sky areas, HST data offers more intricate and detailed insights. The capability to process and classify with HST data unlocks a broad scope for potential studies, both in terms of redshift and magnitude. Furthermore, this establishes a connection for potential applications with James Webb Space Telescope (JWST) data, given its similarity. There are already several studies using HST data as a training set to classify JWST observations \citep{huertas2023galaxy, robertson2023morpheus}.

Initially, the primary objective was to construct a catalog of galaxies observed by both the SDSS (hereafter, SDSS-restricted (SDSS-res)) and the HST. For optimal resolution, we focused on bright galaxies at low redshift. As an illustration, our sample catalog comprises galaxies characterized by low redshifts and high luminosity. Such a selection criterion was imperative for facilitating a seamless transition between SDSS-res and HST datasets. While our dataset was rigorously curated, it has a limited representation of morphological variations for both galaxy types. As such, observed performance might not be optimal.

The initial step involved cross-matching the catalog by \citet{dominguez2018improving} with the Hubble Source Catalog version 3, utilizing a positional matching on Right Ascension (RA) and Declination (DEC) with a 3-arcsecond radius, selecting any following bands: F606W, F775W, or F814W. After removing duplicate entries, non-galactic objects, and objects near the plate boundary, the catalog was refined to 5,438 galaxies. These were then classified, based on \citet{dominguez2018improving}, into Spiral and Elliptical categories based on their \texttt{T-Type} values; specifically, galaxies with \( \texttt{T-Type} > 1 \) and \( \texttt{T-Type} < 0 \) were segregated into separate lists. To augment these lists, galaxies with classifications from the Galaxy Zoo 1 project \citep{lintott2011galaxy}, exhibiting greater than 70\% confidence, were included. Subsequent segmentation was conducted on both the SDSS-res and HST images. It is crucial to note that while these labels were initially generated based on SDSS-res imagery, the low redshift and high luminosity criteria for galaxy selection minimize the likelihood of morphological variation when transitioning to HST data. Finally, we visually inspected each image to ensure there were no fringe cases that could compromise the method's application. This methodology culminated in a refined catalog comprising 654 galaxies, each possessing dual telescope imagery and \texttt{EGG} metric assessments. 
Figure~\ref{fig:sdss_vs_hst_fits} presents the redshift, magnitude, \texttt{T-Type}, and Elliptical-to-Spiral ratio characteristics of the final HST dataset.

The test dataset for each galaxy encompasses imagery from both the SDSS-res and HST, thus enabling a direct comparative analysis of methodological efficacy and metric consistency across both telescopic platforms, independent of data variations. Panels (a) and (b) of Figure~\ref{fig:sdss_vs_hst_ds_fits} serve as exemplars, illustrating the same galaxy as captured by SDSS and HST telescopes respectively.

\begin{figure}
\includegraphics[width=\columnwidth]{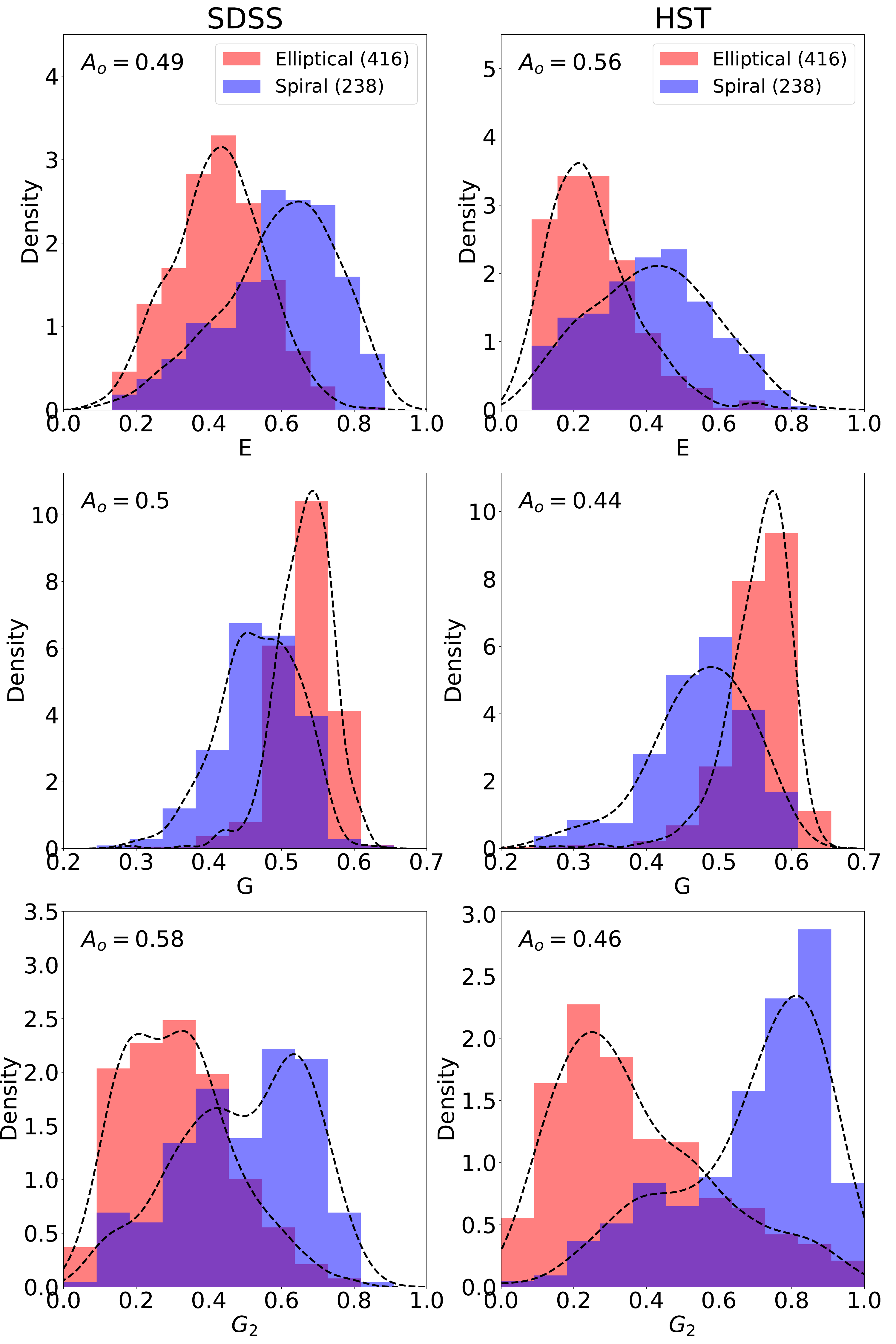}
    \caption{Contrastive analysis of \texttt{EGG} system metrics between SDSS-res and HST datasets. The left column presents results from SDSS-res imagery, while the right from HST. Each row focuses on an individual metric: \texttt{E}, \texttt{G}, and \(\texttt{G}_\texttt{2}\).}
    \label{fig:sdss_vs_hst_v1}
\end{figure}

\subsection{Data Compilation and Preprocessing} \label{hst_data}

The data preparation for the SDSS-res component of our study was straightforward, consisting mainly of FITS cutouts in the r-band. In contrast, preparing the HST data necessitated a more nuanced approach to take into account that, while SDSS is a Survey with wide coverage of the sky, HST is a point to source telescope. In order to maximize our sample size with observations in both experiments, we select filters from the HST that are comparable to the r and i filters of the SDSS. We tested how expanding our analysis to the i filter range can change the morphometrics of the selected galaxies, but we find that both CAS and EGG parameters are mainly unaffected by this selection. Therefore, we select the HST filters F606W, F775W, or F814W, which have central wavelength 5962.20\AA, 7763.11\AA, and 8102.91\AA, respectively, in agreement with the central wavelengths of the r (6231\AA) and i (7625\AA) bands in the SDSS-res. HST data fitting our criteria is retrieved from the Mikulski Archive for Space Telescopes (MAST\footnote{Available in \textit{https://archive.stsci.edu/hst/hsc/}}). Querying among all the available data products, we selected the observation in which the galaxy we are interested in is the further away from the field edges, ensuring that we do not compromise either of the previous described steps (background subtraction, cleaning and segmentation).
To create \textit{RGB} images suited for deep learning algorithms—given that the FITS format is not generally supported—having imagery across a minimum of three filters (Red, Green, and Blue) is essential. While this is not an issue for SDSS-res data, where galaxies are accessible in multiple bands, the limited nature of HST data makes it challenging to compile a significant dataset featuring three color bands for every galaxy. Hence, we converted single-filter data to the \textit{.png} format. Adjustments to stretch and contrast were done using the Trilogy package \citep{coe2012clash}.

\begin{figure*}
    \includegraphics[width=505px]{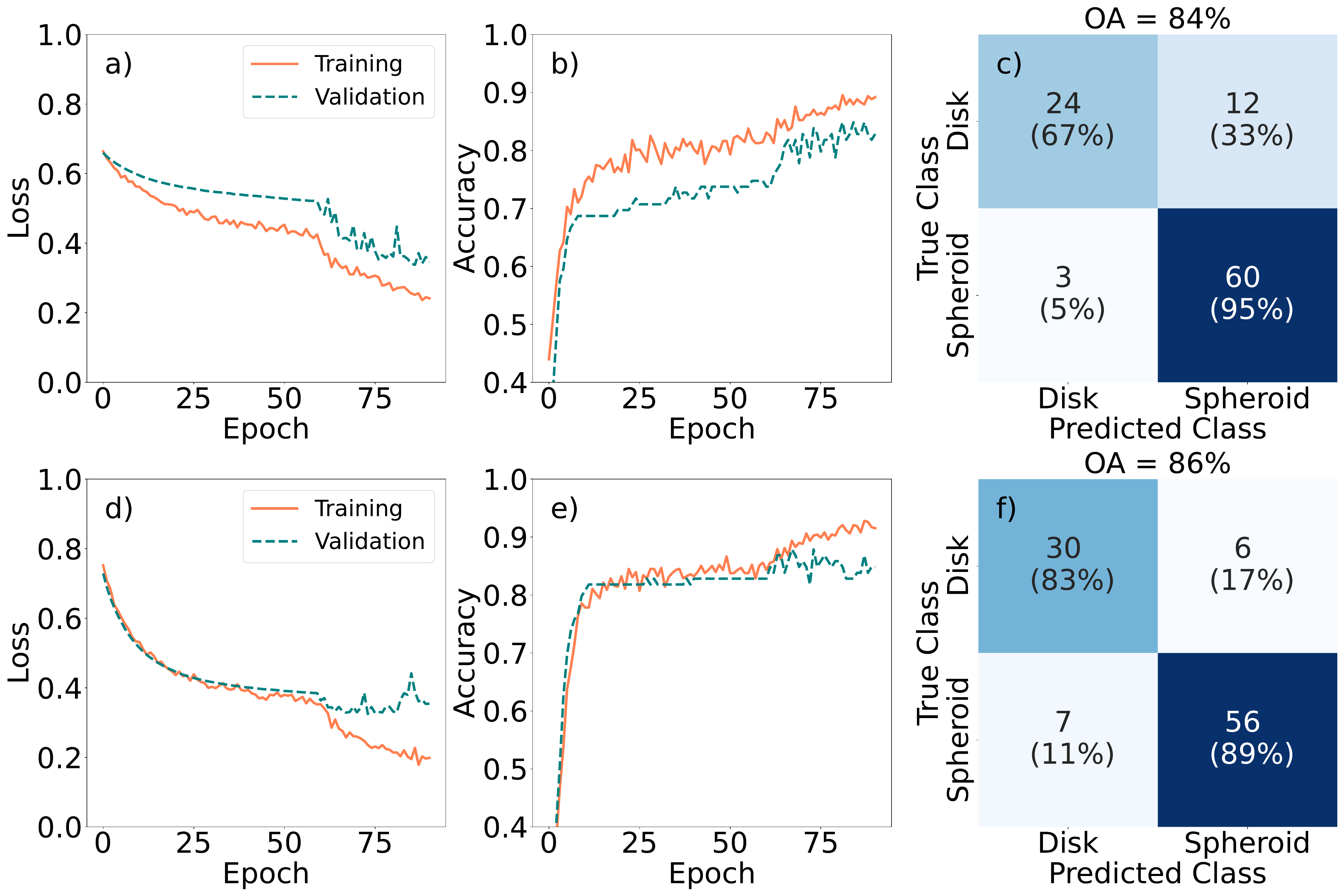}
    \caption{Baseline comparative analysis using SDSS-res (top row) and HST (bottom row) datasets. Each column presents Loss, Accuracy, and the Confusion Matrix based on the imaging data's source.}
    \label{fig:baseline_comparison}
\end{figure*}

For effective segmentation and metric extraction from HST data, we had to diverge from the standard pipeline outlined in Section~\ref{method}. A primary alteration was enhancing the image preprocessing phase, which necessitated downsampling to account for the more pronounced details in HST images compared to SDSS-res. These intricate details, like star-forming regions and fluctuations in gas density, encapsulate a wealth of information that can potentially induce metric instability, notably in the context of \(\texttt{G}_\texttt{2}\). From our empirical studies, we found that the gradient field is highly vulnerable to perturbations, causing instability. 
It's crucial to note that metrics were devised and serve as numerical descriptors capturing general morphological features of galaxies, such as the disk, bulge, spiral arms, and the flux distribution across the galactic surface. These features suffice for confidently distinguishing between the primary morphological classifications. Conversely, examining data replete with high-frequency details and probing into subtler aspects like star-forming regions, we begin to challenge the ground definition of metrics. Moreover, this excess of high-frequency information might overshadow the general morphological traits, jeopardizing metric performance.
A visual demonstration of this behavior, the differences in level of details and the perturbations in the gradient field can be viewed in Figure~\ref{fig:sdss_vs_hst_ds_fits}. Furthermore, the downsampled version (panel f) accentuates the visual features of spiral galaxy, that were not visually noticeable in original HST image (panel e). The downsampling was applied solely to the data used for metric extraction. No additional manipulations were made to the data used for DL since a CNN model is anticipated to leverage all the details for learning. Following this reasoning, downsampling could potentially degrade the performance of the CNN.

The downsampling process utilized the \texttt{block\_reduce} function from the SciPy library. Specifically, the function was called as:
\begin{align*}
\texttt{image} &= \texttt{block\_reduce} \\
&\quad \texttt{(image, block\_size=(factor, factor),} \\
&\quad \texttt{func=np.mean)}.
\end{align*}
Here, \( \texttt{factor} \) is determined as \( \texttt{factor = pixel\_scale\_sdss / pixel\_scale\_hst} \). It is essential to account for different instruments and detectors potentially having unique pixel scales. The results of this downsampling can be viewed in Panel (c) of Figure~\ref{fig:sdss_vs_hst_ds_fits}, where the downsampled HST image matches the SDSS-res image dimensions but retains enhanced detail. 

After implementing these modifications, we conducted a comparative analysis of metric performance across SDSS-res and HST datasets. This concludes selection of the data for SDSS-res versus HST comparison. Resulting selection, we call demonstration catalog.

\subsection{Comparative Morphological Analysis}

Figure~\ref{fig:sdss_vs_hst_v1} showcases the metric-based differentiation of galaxy morphologies using the \texttt{EGG} system across both datasets. Galaxies were processed twice: once with SDSS-res and once with HST data.

Metrics consistently demonstrated their ability to differentiate galaxies in both datasets. For metrics like \texttt{G} and \(\texttt{G}_\texttt{2}\), the HST dataset seemed to enhance the distinction between galaxy classes compared to the SDSS-res dataset.

This data was then fed into our unsupervised classification algorithm. To evaluate the HST data's impact on our methodology, we ran the complete analysis pipeline for each dataset, offering a thorough performance comparison.

\subsection{Deep Learning Baseline}

To robustly validate the efficacy of our proposed unsupervised method, it is essential to first establish a performance benchmark for reference. Our chosen benchmark is a supervised deep learning model, specifically the Xception architecture, which was detailed in section~\ref{method}. This model, trained on labeled galaxy morphologies from our catalog, provides a heuristic indicator of what performance might be achievable with unsupervised techniques. While Section~\ref{method} did not necessitate such a baseline — primarily because an OA above 99\% was reasonably expected given the exhaustive representation of galaxy morphologies in the test catalog — the demonstration catalog does not enjoy this comprehensive coverage. Some galaxy morphologies are underrepresented or missing, thus underscoring the importance of a supervised performance baseline.

Figure~\ref{fig:baseline_comparison} provides an overview of the Xception model's performance when trained on the demonstration catalog. It is commendable that the model achieved reasonable performance levels even with limited data. For this test, the data was split into training (85\%) and validation (15\%) segments, with the lack of a dedicated testing dataset attributable to data constraints. As outlined in Section~\ref{method}, the training process leveraged both transfer learning and data augmentation strategies. Panels c and f of Figure~\ref{fig:baseline_comparison} showcase the confusion matrices from the validation data. Expectedly, the HST dataset yielded a superior OA and better-balanced misclassification rates compared to the SDSS-res dataset. This could likely be due to the richer details present in the HST data; it is worth noting that no downsampling was employed for the deep learning model.

This baseline performance serves as a touchstone when assessing the results from our unsupervised method.

\begin{figure}
    \includegraphics[width=\columnwidth]{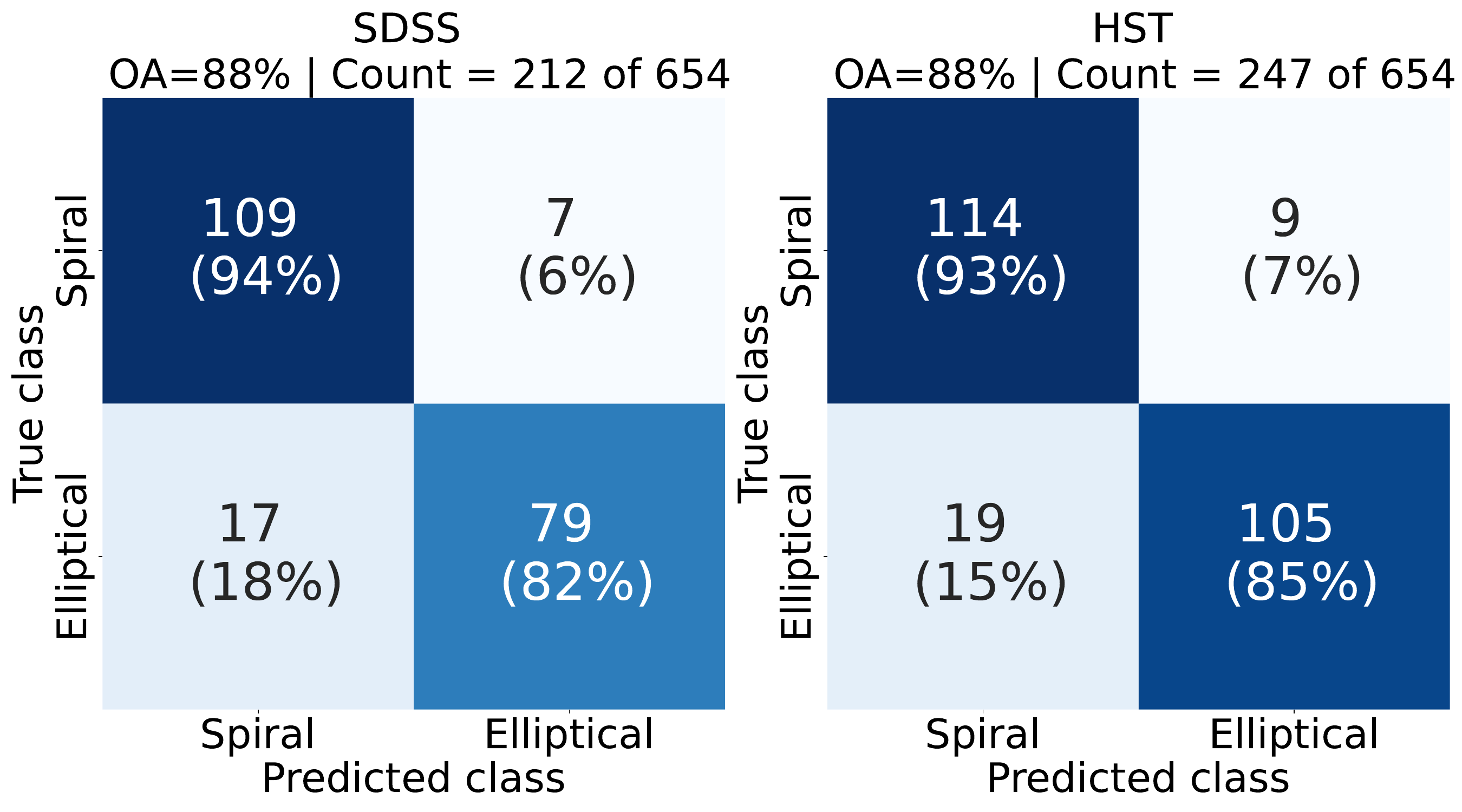}
    \caption{Confusion matrices representing the selection from prominent clusters for SDSS-res (left) and HST (right) datasets. These matrices capture misclassifications while also enumerating the galaxies chosen from the total available.}
    \label{fig:prominent_sdss_vs_hst}
\end{figure}

\subsection{Application and Evaluation of the Unsupervised Method}
\label{sec:method-application}

In this subsection, we detail the application of our proposed unsupervised methodology to both SDSS-res and HST datasets. The overarching goal is to gauge the method's adaptability and efficacy, especially with high-resolution telescope data like that from the HST. A thorough exposition of the method's foundations can be found in Section~\ref{method}. Here, we focus on modifications induced by data constraints and delve into the consequent outcomes.

Two core metrics gauge the method's success: (1) the accuracy of prominent clusters selected from the unsupervised procedure, and (2) the performance of a Convolutional Neural Network (CNN) when trained on labels deduced from these clusters and tested on new data. However, while Section~\ref{method} had an expansive dataset facilitating a rigorous appraisal, our demonstration catalog offers only 654 samples, compelling us to devise specific adaptations. Given the pressing challenge of limited data, we have embraced the following data allocation strategy:

\begin{itemize}
    \item All catalog entries undergo unsupervised clustering.
    \item Galaxies identified within prominent clusters constitute the CNN model's training data.
    \item Remaining galaxies, those outside prominent clusters, function as the test set for the CNN evaluation.
\end{itemize}

\begin{figure}
    \includegraphics[width=\columnwidth]{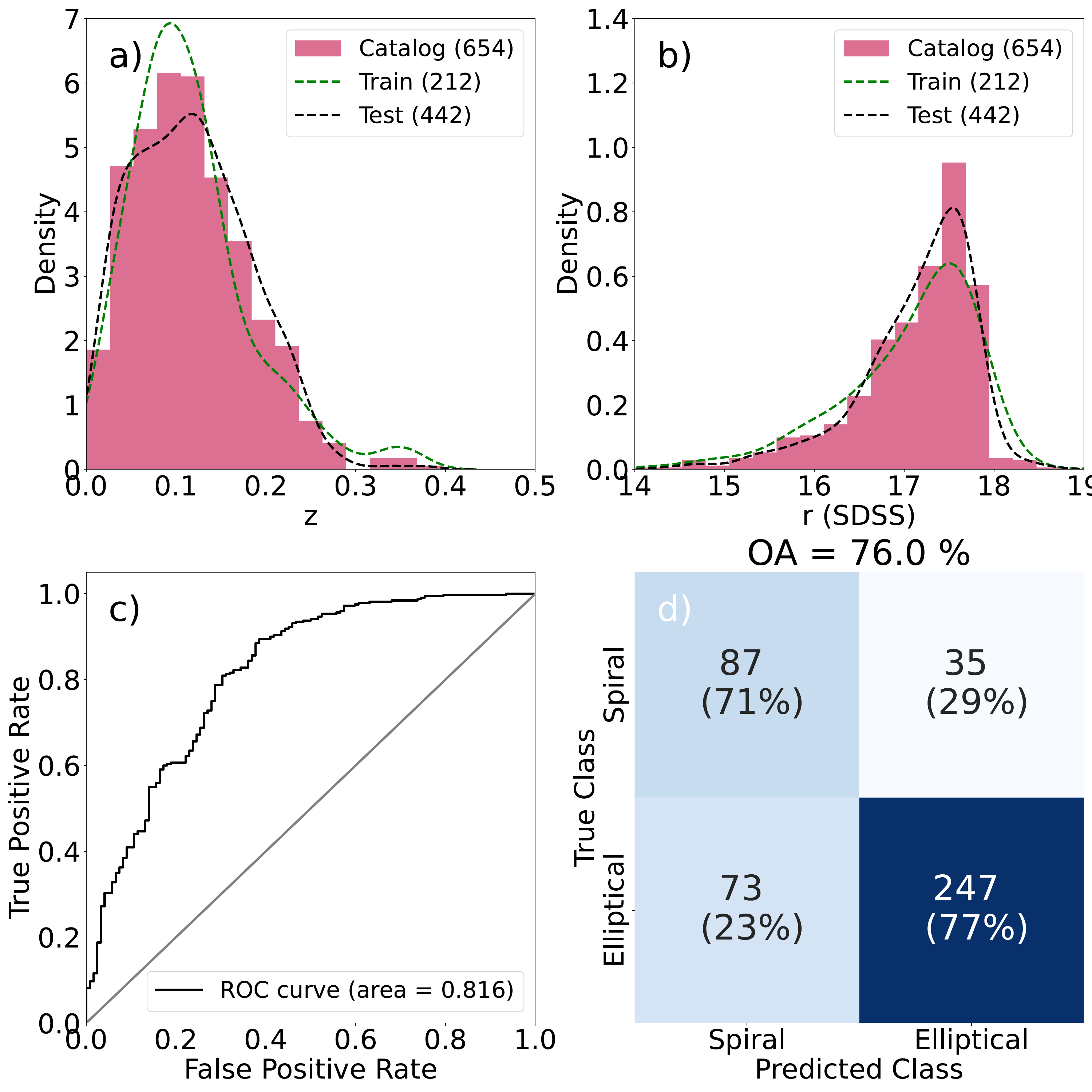}
    \caption{Evaluative view of the CNN model's performance when trained using unsupervised labels for the SDSS-res dataset. Panels (a) and (b) juxtapose the complete catalog with the subset designated for testing. Panels (c) and (d) exhibit the ROC curve and respective confusion matrix for the testing segment.}
    \label{fig:test_not_prominent_sdss_v1}
\end{figure}

This methodology
%
provides a balanced assessment within the confines of our dataset. It ensures that the CNN undergoes training and testing on distinct galaxy groups, but with analogous redshift and magnitude spectrums. It is pivotal to note that for the testing phase, we retain the original catalog labels for galaxies, sidestepping labels from the unsupervised clustering. Only galaxies clustered in prominent clusters are labeled according to their clustering position, serving as the supervised dataset for CNN training. Such an approach guarantees uniform experimental conditions across both SDSS-res and HST datasets, preserving the method's integrity.

\subsection{Analysis of Method Application}
\label{sec:analysis-method-application}

We initiated our assessment with the identification and selection of prominent clusters. As depicted in Figure~\ref{fig:prominent_sdss_vs_hst}, the confusion matrices for both SDSS-res and HST datasets post-cluster selection reveal compelling insights. Aligning with our earlier observations from Section~\ref{method}, we noted that $2/3$ to $1/2$ of the galaxies are excluded after focusing on prominent clusters in both datasets. The OA is largely congruent with baseline evaluations, with the HST data demonstrating a slight edge in class-specific accuracy metrics over the SDSS-res data.

Post the cluster selection phase, we divided the catalog into dedicated training and test segments. Those galaxies identified within the prominent clusters (as highlighted in Figure~\ref{fig:prominent_sdss_vs_hst}) constituted the training pool for the CNN model, while the rest were marked for testing. Throughout the training, we relied on labels deduced from unsupervised clustering, while the test phase made use of the original catalog labels.

\begin{figure}
    \includegraphics[width=\columnwidth]{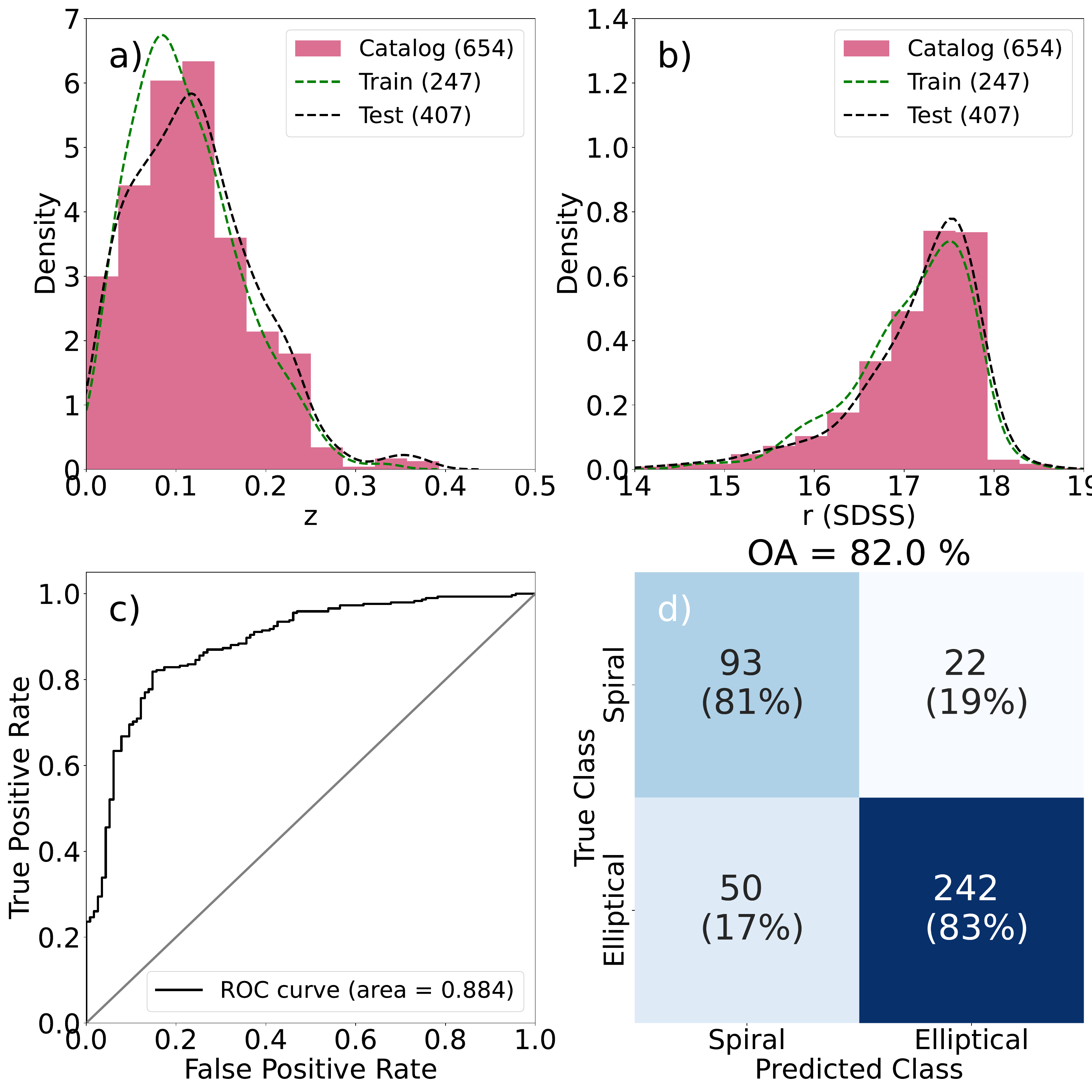}
    \caption{Performance assessment of CNN model trained on unsupervised labels for HST data. Panels (a) and (b) contrast the overall catalog data with the subset used for CNN testing. Panels (c) and (d) present the ROC curve and confusion matrix for the test set.}
    \label{fig:test_not_prominent_v1_hst_v1}
\end{figure}

\subsubsection{Results analysis}
Detailed insights into the SDSS-res dataset are provided in Figure~\ref{fig:test_not_prominent_sdss_v1}. Panels (a) and (b) elucidate the distribution of the full catalog against its testing subset, underscoring the parallels between them. Panels (c) and (d) delve into the performance of our CNN model, which, unfortunately, does not fare optimally. This shortcoming can be attributed to the circumscribed diversity and quantity of galaxy samples within our dataset made available for CNN training.

When applied to the HST dataset, our method's outcomes were consistent but showcased an improved performance. As revealed in Figure~\ref{fig:test_not_prominent_v1_hst_v1}, the model trained with HST data clearly outstripped its SDSS-res equivalent, registering advancements in both OA and class-wise accuracy. This progression underscores the assertion that superior quality, detail and resolution wise, input data can indeed catalyze model performance enhancements. In sum, the deployment of our method on HST data corroborates that a leap in input data quality can significantly elevate model performance, particularly by leveling misclassification rates.

However, it it worth noting that these results should be taken with caution, given the fact the original catalog does contain multiple subjects that have have classification ambiguity and those introduce complexity in the model training and classification. In a case of bigger catalog, we could filter it similarly to the Section \ref{ttype_bias}. This would improve OA and general model behavior.

\section{Summary}\label{section_conclusions}

We present a novel approach to classifying galaxies into different morphological types. The method shows promising results in discerning classes and constructing a reliably labeled catalog. This catalog can be used as input to supervised CNNs to classify much larger datasets using the trained model, significantly enhancing its scalability. The approach is grounded on a newly compiled morphological system called \texttt{EGG}, comprising Entropy, Gini, and Gradient Pattern Analysis. In Figures \ref{fig:cas_vs_egg} and \ref{fig:cas_vs_egg_ttype}, we present a comparison, through the lens of morphological separation, between the \texttt{EGG} and \texttt{CAS} systems. We argue that \texttt{EGG} shows the better separation, justifying the choice of the system as a basis for unsupervised clustering.

To summarize, the approach undertaken in the processing pipeline of the present method elucidates a few critical outcomes:

\begin{itemize}
    \item \textit{Foundation in Physical Reality:} Being rooted in morphological metrics, the methodology stands on a foundation that is conceptually robust from a physical perspective. The morphological features of galaxies, which are intrinsically tied to their formation and evolution processes, offer a scientifically grounded criterion for classification.

    \item \textit{High unsupervised accuracy:} Comparing our OA results from prominent cluster selection with Table 3 in \citet{barchi2020machine}, we note that the purely unsupervised method rivals Traditional Machine Learning methods such as Decision Tree (DT), Support Vector Machine (SVM), and Multilayer Perceptron (MLP). On a similar, yet more complex dataset, we achieve approximately 95\% of OA, while DT, SVM, and MLP oscillate between 94\% and 95\%.
    
    \item \textit{Alignment with Supervised Methods:} The unsupervised morphological classification approach has showcased its capability to reproduce results analogous to those derived from a supervised model. In other words, the supervised model agrees with the unsupervised labeling and is able to learn patterns based on this classification. This alignment implies that there is substantial overlap between the knowledge derived from labeled data and the intrinsic structures captured by unsupervised learning.
    
    \item \textit{Efficiency and Scalability:} The developed method allows for the rapid and automated creation of labeled catalogs. This efficiency not only reduces the manual labor and expertise required for labeling but also offers scalability for large datasets.
    
    \item \textit{Model Reusability:} The resultant Convolutional Neural Network (CNN) model, exhibiting high accuracy, can be seamlessly applied to datasets bearing similarity, without necessitating retraining. This adaptability is instrumental in facilitating broader applications in the field of galaxy morphology.
\end{itemize}

In essence, the synergy of unsupervised learning techniques with the foundational principles of astrophysics fosters a comprehensive and robust solution for galaxy classification challenges. Furthermore, we analyzed the application of our proposed unsupervised method for galaxy classification using both Sloan Digital Sky Survey (SDSS) and Hubble Space Telescope (HST) data. The findings offer compelling evidence of the method's efficacy, substantiated by comparison metrics, deep learning baselines, and real-world application scenarios. Firstly, our analysis showed that the morphological metrics used are robust across different data sources, namely SDSS and HST, with some metrics even showing improved separation in higher-quality HST data. This underscores the versatility of the \texttt{EGG} system in handling data from different surveys. Secondly, a baseline performance comparison using a supervised deep learning approach revealed a high degree of accuracy, setting a challenging standard for our unsupervised method. Remarkably, the unsupervised method was able to approach, and in some instances exceed, this high benchmark—particularly when applied to HST data. Thirdly, despite limitations in data volume and diversity, the method demonstrated commendable adaptability. It achieved high Overall Accuracy and class-specific accuracy, indicating that the quality of input data is a significant factor for the method's success. Importantly, the HST data showed a more balanced misclassification rate, which not only underscores the potential for applying the method to more advanced telescopes and datasets, but also improves the reliability of the galaxy classifications. However, it is crucial to remember that the demonstration catalog is inherently limited by its focus on low-redshift and bright, well-resolved galaxies. Further studies would be needed to validate the method's generalizability across a broader range of galaxy types and observational conditions. In summary, the results validate the utility and robustness of the proposed unsupervised method for galaxy classification, and they bode well for its future application in more extensive and diverse datasets. 

\section{Acknowledgements}
This work was financed in part by the Coordenação de Aperfeiçoamento de Pessoal de Nível Superior – Brazil (CAPES) – Finance Code 001. VMS and RRdC acknowledge the support from FAPESP through the grants 2020/15245-2 and 2020/16243-3, respectively.
\section*{Data Availability}
The data that support the findings of this study are available from the corresponding author, IK, upon reasonable request.
 



\bibliographystyle{mnras}
\bibliography{ref} 


\appendix
\newpage

\section{HST data preprocessing - high redshift friendly approach}\label{sec:appendix2}
\begin{figure}
	\includegraphics[width=\columnwidth]{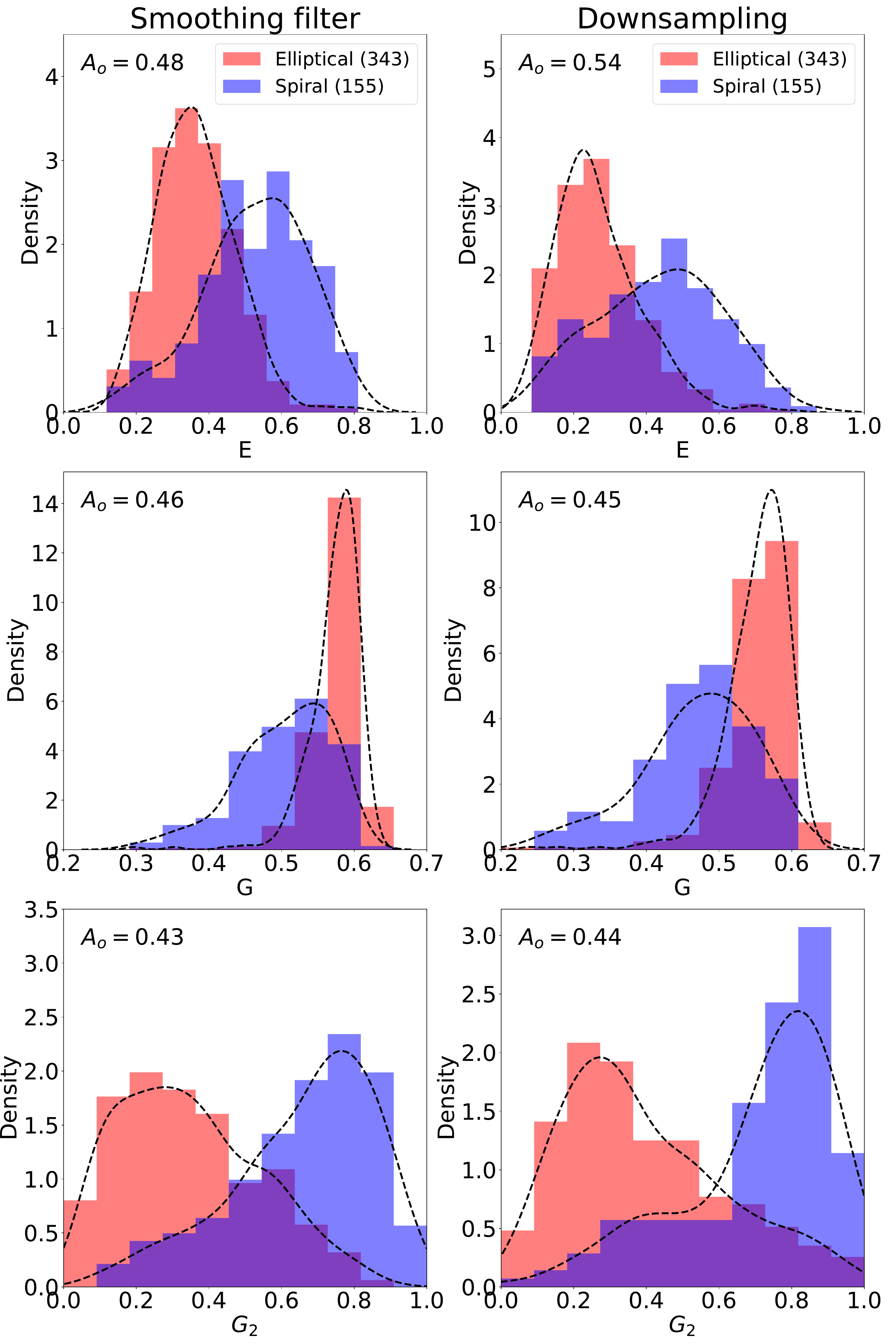}
    \caption{Comparison between prepossessing techniques for HST images preprocessing. First columns shows results for Smoothing and second column shows the results for Downsampling. $A_o$ provides the values of overlap of distributions.}
    \label{fig:epanechnikov_vs_downsample_v1}
\end{figure}

In section \ref{section_sdss_hst} we present an application of the method on the HST data. To optimize the metric extraction from the HST data, it is necessary to apply preprocessing to the image beforehand. This preprocessing is required to decrease the level of details that are found in the HST images. Through empirical exploration, we discovered that an excessive amount of details in HST images could be detrimental to metric extraction. This forms the motivation to apply preprocessing.

Across all tested preprocessing methods, we single out two as the most effective: Downsampling and Smoothing with the Epanechnikov filter. In Figure \ref{fig:epanechnikov_vs_downsample_v1}, we show the comparison between the results obtained by applying both preprocessing techniques. Downsampling involves applying the \texttt{block\_reduce} function to the HST image, aiming to decrease its resolution to one similar to SDSS. This results in a combination of blocks of pixels and a consequent decrease in the total amount of information in the image. More detail can be found in \ref{hst_data}. Smoothing the image involves applying the Epanechnikov filter to the image to smooth out the excess of flux variation while preserving the same size of the image and the galaxy therein. Both of these filters provide similar performance, with a slight edge for Downsampling in terms of metric separation and processing speed (the output image is smaller and processed faster). In this research, based on better immediate performance, both in separation and speed, we decided to use Downsampling. However, it is important to note a detail of image degradation. Given the fact that we operate in low-redshift, where all of our data is well resolved and bright, we do not run into the risk of losing too much of the details when Downsampling. The problems could arise when moving to higher redshifts, with data that will naturally contain smaller galaxies. In such cases, Downsampling could remove too much detail, along with the galaxy structure. That is why, potentially, it would be wise to further investigate smoothing the images,
yet preserving the main features of the galaxy.

\section{Unsupervised Classification - Alternative Approaches}\label{sec:appendix1}

\begin{figure}
	\includegraphics[width=\columnwidth]{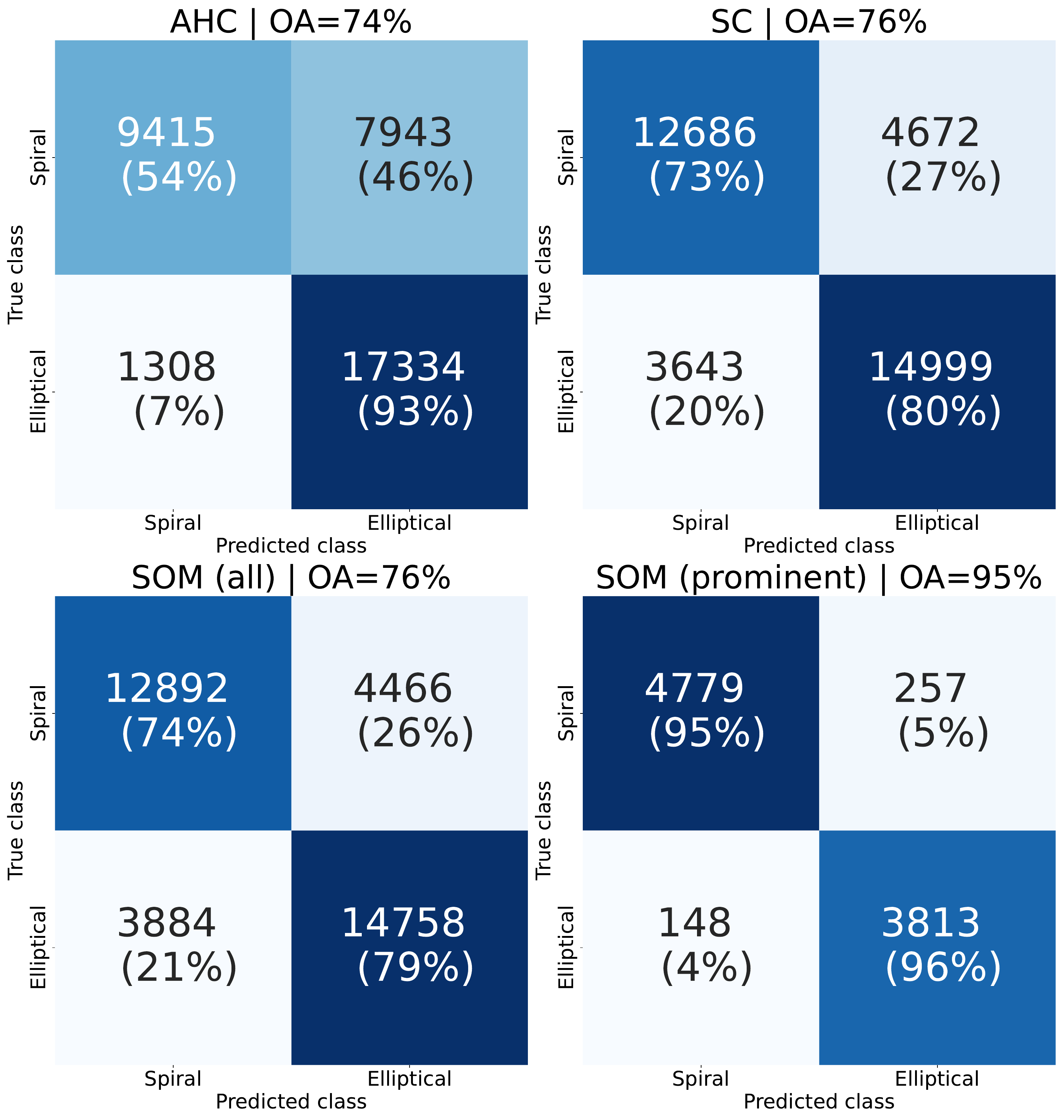}
    \caption{Unsupervised classification comparison for the four tested methods. Prominent clusters were used for SOMbrero.}
    \label{fig:alt_unsup_comparison}
\end{figure}

\begin{figure*}
	\includegraphics[width=505px]{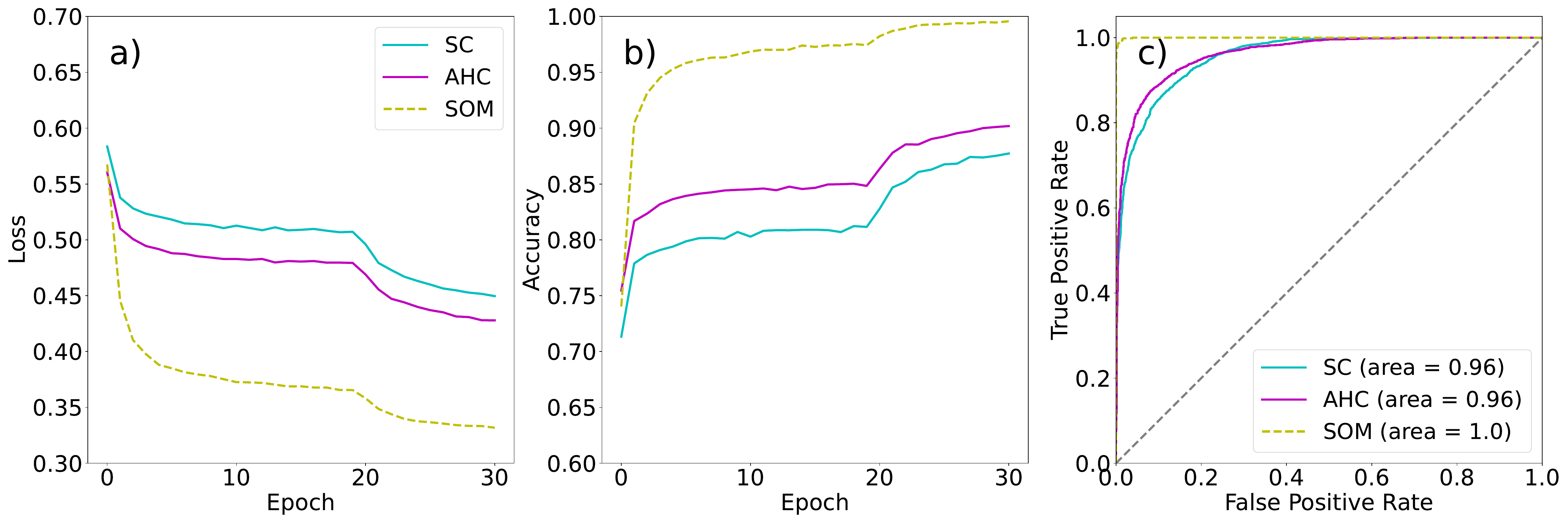}
    \caption{Training report for alternative clustering algorithms. Panel (a) displays the loss curves, panel (b) illustrates the accuracy, and panel (c) presents the ROC curves based on the ensemble results of the validation catalog.}
    \label{fig:alt_training}
\end{figure*}

We present an analysis comprising two alternative techniques—all of which are unsupervised clustering algorithms operating on tabular data constituted of morphological metrics. These two methods, namely Agglomerative Hierarchical Clustering and Spectral Clustering, demonstrated strong performance among a greater number of tested approaches. Our goal here is to test the how reliable is SOMbrero as our main unsupervised algorithm via comparison with several alternatives. We perform similar (adapted where needed) steps to provide proof that the methodology shows consistent results across the algorithms. Of course, SOMbrero will have an edge, primarily due to its robust functional base, documentation, and greater flexibility. Moreover, being based on Self-Organizing Maps, it enables functionality inaccessible to simple clustering algorithms. For each method, we will use the same input (a file containing morphological metrics) as was used with SOMbrero. Each method is tuned to its optimal parameters and provides a predicted label based on its clustering routine. After the clustering step, we use the output (predicted labels) to train a supervised CNN and evaluate it first on a test portion of the training data, then on the remainder of the \citet{dominguez2018improving} and GZ1 \citep{lintott2011galaxy} catalogs defined in Section \ref{sec:data}. For each algorithm test, we used the same training and testing parameters, including the number of epochs, ensemble models, and the level of label smoothing for noise management as was used with SOMbrero.

\subsection{Agglomerative Hierarchical Clustering (AHC)}
Agglomerative Hierarchical Clustering employs a bottom-up clustering strategy. Initially, each individual data point is treated as a standalone cluster. These are then progressively merged based on a specific similarity measure until only a single cluster remains. This process yields a tree-like diagram known as a dendrogram, which elucidates the hierarchical relationships between clusters. A significant advantage of this method is the ability to choose the number of clusters based on the required granularity. However, a limitation is that once clusters have been merged, the process is irreversible, potentially leading to premature commitment.

\subsection{Spectral Clustering (SC)}
Spectral Clustering leverages the principles of graph theory and linear algebra to group similar data points into clusters. It constructs a similarity graph from the data, where nodes represent data points and edge weights denote the similarity between pairs of these points. By using eigenvectors derived from the graph's Laplacian, the method reduces the dimensionality of the data. It then applies a clustering technique (such as K-means) within this reduced-dimensional space. Spectral Clustering is particularly effective for datasets with non-globular structures, allowing it to identify clusters of varying shapes and sizes.

\subsection{Comparison}

We begin our comparison with unsupervised classification. Figure \ref{fig:alt_unsup_comparison} displays the confusion matrices, showcasing the results of how each method classified an unsupervised catalog of 36000 based on morphological metrics. For our primary clustering algorithm, SOMbrero, we show total and prominent cluster selections. It is evident that both alternative methods yield a similar classification score, hovering around 75\%. A point worth noting is that Spectral Clustering (SC) demonstrates superior per-class misclassification, while AHC tends to misclassify Spiral galaxies significantly, almost bordering guess probability. On the other hand, SOMbrero, while using all clusters, shows similar performance to SC. However, switching to only prominent cluster, it shows a clear advantage over two algorithms, boasting 95\% OA and a lower and balanced per-class misclassification rate.

\begin{figure*}
	\includegraphics[width=505px]{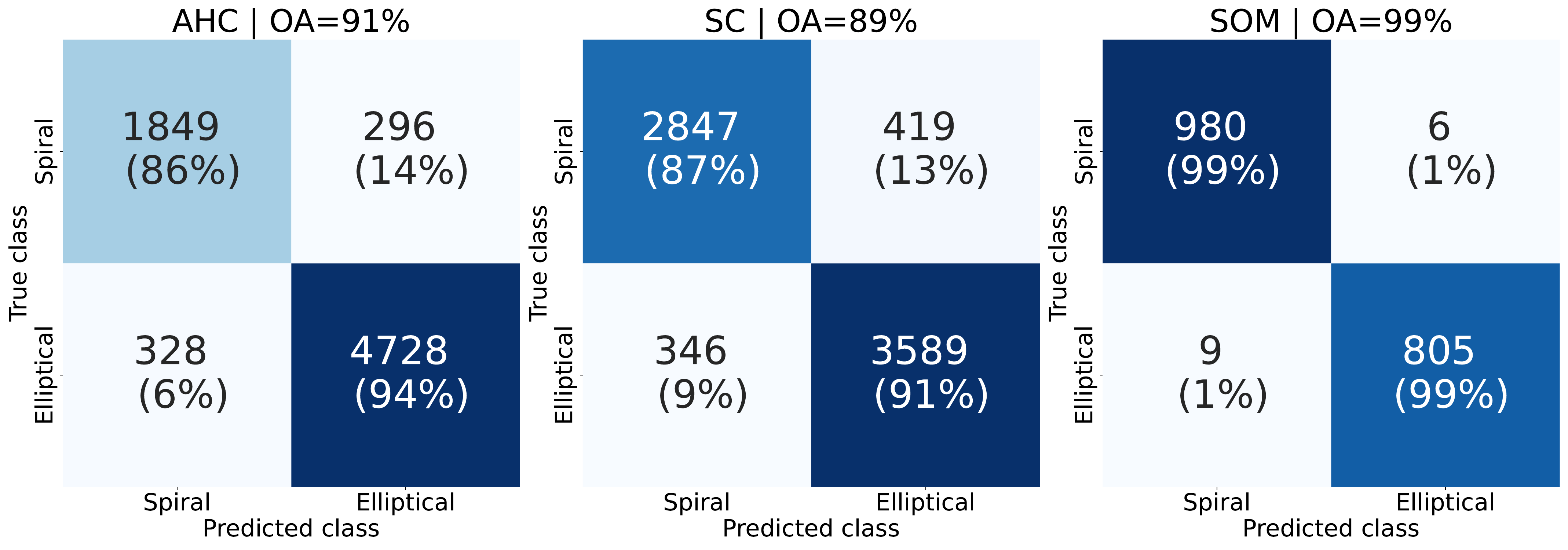}
    \caption{Resulting confusion matrices after CNN training and application of the ensemble on the validation catalog.}
    \label{fig:cm_sup_alt}
\end{figure*}

The next step in our pipeline involves assigning the labels obtained from unsupervised clustering and using the data to train a CNN model with those labels. Figure \ref{fig:alt_training} depicts the training behavior of the CNN model when utilizing data from each of the algorithms. For all three methods, a consistent behavior is observed, characterized by a gradual decrease in loss (as shown in panel a) and an increase in accuracy (depicted in panel b). However, despite the general trends being similar, SOMbrero clearly outperforms in both metrics, underscoring that internal agreement and reduced noise are vital for successful CNN training. Panel (c) in Figure \ref{fig:alt_training} presents the ROC curves based on the ensemble results of the validation catalog. Yet again, AHC and SC exhibit comparable behavior, while SOMbrero stands out with its superior model performance.

In Figure \ref{fig:cm_sup_alt}, we delve into a more detailed comparison of the CNN models, each trained with corresponding input labels generated by their respective clustering algorithms. The trends observed in previous plots persist: AHC and SC exhibit similarities in behavior, while SOMbrero remains notably superior. The two alternative methods yield relatively commendable results, almost attaining a 90\% OA. However, per-class misclassifications are still significant, especially for Spiral galaxies, when compared to SOMbrero.

\begin{figure*}
	\includegraphics[width=505px]{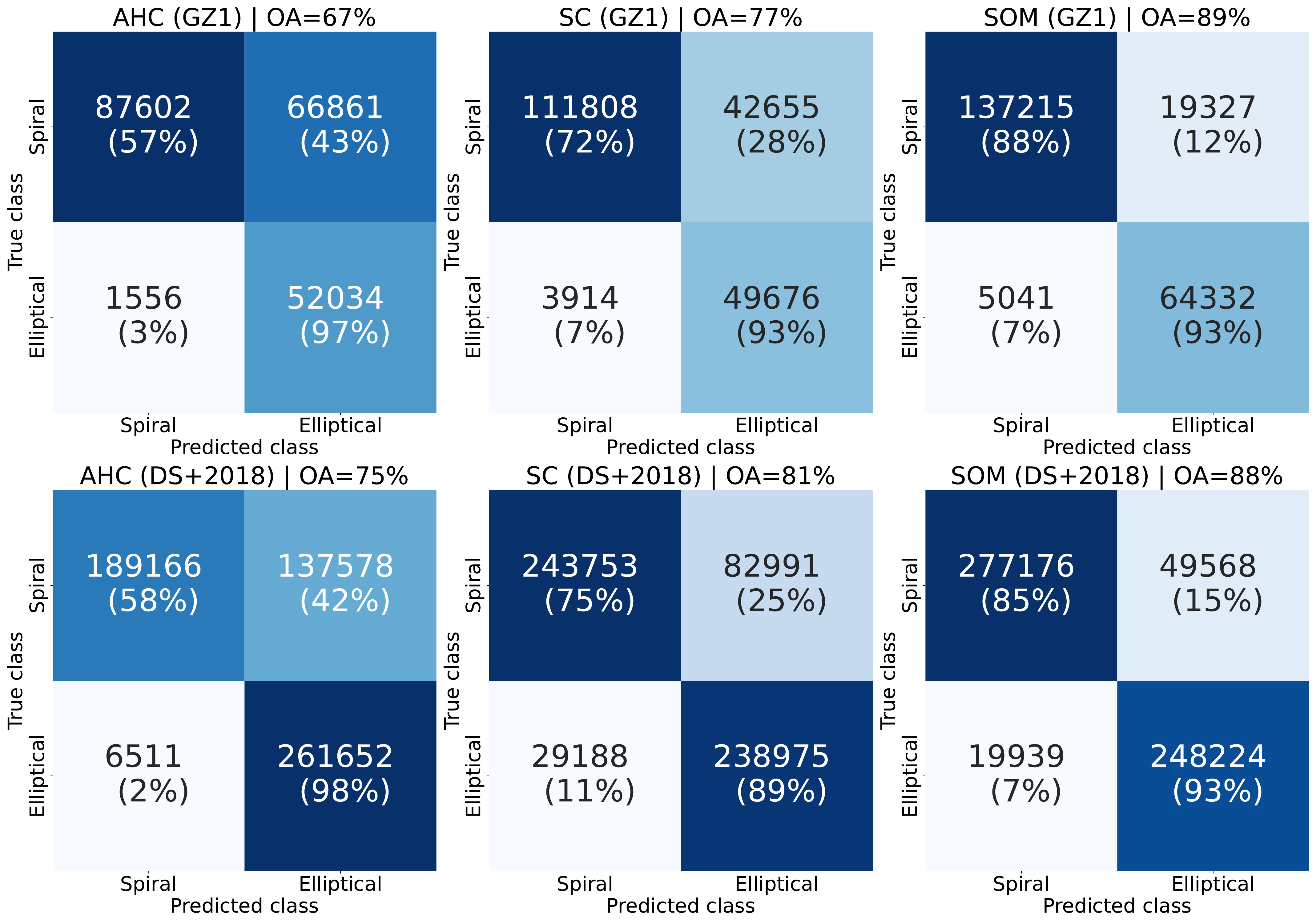}
    \caption{Resulting confusion matrices after CNN training and application of the ensemble on the test catalogs.}
    \label{fig:cm_sup_test_alt}
\end{figure*}

Lastly, in Figure \ref{fig:cm_sup_test_alt}, we illustrate the confusion matrices for each of the two primary test catalogs (filtered from galaxies in range $-1 < \texttt{T-Type} < 0$), contrasting the performance of CNNs trained on labels produced by each algorithm. Notably, during this concluding step, we note different performance in all three tested algorithms, with  SOMbrero on top.  Comparatively, all three methods offer distinct approaches to clustering. AHC is optimal when a hierarchy of clusters is essential, while SC excels at recognizing complex cluster shapes. In the context of galaxy classification, all three algorithms rendered comparable performance, with SC consistently delivering the highest accuracy across multiple test datasets. 

Each method grapples with a similar challenge: a marked decline in performance when classifying Spiral galaxies and a high noise level for both classes. Given such pronounced noise, the supervised model finds it challenging to discern and assimilate the distinguishing features of each class. The repercussions of this are evident during the training step (as shown in Figure \ref{fig:alt_training}), where, in general, models exhibit elevated loss and diminished OA. This adversely affects the eventual model testing outcomes as depicted in Figures \ref{fig:cm_sup_alt} and \ref{fig:cm_sup_test_alt}. Consequently, it is inferred that CNNs could not identify patterns in the data as effectively as with SOMbrero, where, during the training phase, we achieved an OA exceeding 98\%. Even though the OA surpasses the chance level (50\%), all models trained using alternative methods grapple with class differentiation.

Contrasting these with our primary algorithm, the SOMbrero R package, several distinctions become evident. SOMbrero employs Self-Organizing Maps (SOMs), a type of artificial neural network trained with unsupervised learning to craft a low-dimensional, discretized portrayal of the training samples' input space. Divergent from the clustering methods previously mentioned, SOMbrero can project high-dimensional data onto a lesser-dimensional realm, preserving both topological and metric relationships. This characteristic renders it especially adept at visualizing high-dimensional data and pinpointing non-linear connections. Furthermore, the package's formidable visual inspection tools enable superior outcomes with minimal human oversight during cluster selection. This capacity aids in circumventing clusters with extensive class mixtures, streamlining the training of the CNN, and reducing label/class discrepancies.

In summation, each method offers a unique clustering and classification approach, each bearing its strengths and limitations. Nonetheless, our primary method, SOMbrero, furnishes a superior environment and capability for unsupervised galaxy classification, culminating in an enhanced overall accuracy. The foundational accuracy, immediately post the clustering phase (considering SOMbrero's full grid), reveals analogous OA across all four algorithms. In essence, SOMbrero does not initially eclipse the alternatives. Only subsequent to the prominent cluster selection does SOMbrero distinctly surpass its counterparts, delivering an output precise enough to act as the ground truth for CNN training. This comparison's pivotal goal was to highlight that the outcomes procured are not solely contingent on a specific algorithm and that relative uniformity is upheld across diverse methodologies.


\bsp	
\label{lastpage}
\end{document}